\documentclass[]{aa}  

\usepackage{graphicx}
\usepackage{txfonts}
\usepackage{xcolor,colortbl}
\usepackage{hyperref}
\usepackage{multirow}
\usepackage[normalem]{ulem}
\makeatletter
\renewcommand*\aa@pageof{, page \thepage{} of \pageref*{LastPage}}
\makeatother
\newcommand{\teff}{$T_{\rm eff}$}
\newcommand{\logg}{$\log(g)$}
\newcommand{\feh}{[Fe/H]}
\newcommand{\sife}{[Si/Fe]}
\newcommand{\mgfe}{[Mg/Fe]}
\newcommand{\cafe}{[Ca/Fe]}
\newcommand{\vmic}{$\xi_{\rm micro}$}
\newcommand{\vmac}{$\xi_{\rm macro}$}
\newcommand{\noline}{\cellcolor{black!00}-}

\begin{document} 

   \title{Detailed $\alpha$ abundance trends in the inner Galactic bulge}

   \author{N. Nieuwmunster
            \inst{1,2}
            \and
            G. Nandakumar
            \inst{2}
            \and
            E. Spitoni
            \inst{1,3}
            \and
            N. Ryde
            \inst{2}
            \and
            M. Schultheis
            \inst{1}
            \and
            R. M. Rich
            \inst{4}
            \and
            P. S. Barklem
            \inst{5}
            \and
            O. Agertz
            \inst{2}
            \and
            F. Renaud
            \inst{2}
            \and
            F. Matteucci
            \inst{6,7,8}
            }

   \institute{Université Côte d’Azur, Observatoire de la Côte d’Azur, Laboratoire Lagrange, CNRS, Blvd de l’Observatoire, 06304 Nice, France\\
              \email{niels.nieuwmunster@oca.eu}
    \and         
    Lund Observatory, Department of Astronomy and Theoretical Physics, Lund University, Box 43, SE-22100 Lund, Sweden 
    \and Stellar Astrophysics Centre, Department of Physics and
  Astronomy, Aarhus University, Ny Munkegade 120, DK-8000 Aarhus C,
  Denmark  
    \and
    Department of Physics and Astronomy, UCLA, 430 Portola Plaza, Box 951547, Los Angeles, CA 90095-1547 
    \and
    Theoretical Astrophysics, Department of Physics and Astronomy, Uppsala University, Box 516, 751 20 Uppsala, Sweden 
    \and
    Dipartimento di Fisica, Sezione di Astronomia, Università di Trieste, via G.B. Tiepolo 11, I-34131, Trieste, Italy 
    \and
    I.N.A.F. Osservatorio Astronomico di Trieste, via G.B. Tiepolo 11, I-34131, Trieste, Italy 
    \and
    I.N.F.N. Sezione di Trieste, via Valerio 2, 34134 Trieste, Italy 
    }

   \date{Received Month xx, xxxx; accepted Month XX, XXXX}

 
  \abstract
   {Until now,  heavy interstellar extinction has meant that only a few studies of chemical abundances  have been possible in the inner Galactic bulge. However, it is crucial to learn more about this structure in order to better understand the formation and evolution of the centre of the Galaxy and galaxies in general.}
  {In this paper, we aim to derive high-precision $\alpha$-element abundances using CRIRES high-resolution IR spectra of 72 cool M giants of the inner Galactic bulge.}
   {Silicon, magnesium, and calcium abundances were determined by fitting a synthetic spectrum for each star.  We also incorporated recent theoretical data into our spectroscopic analysis (i.e. updated K-band line list, better broadening parameter estimation, non-local thermodynamic equilibrium (NLTE) corrections).  We compare these inner bulge $\alpha$ abundance trends with those of solar neighbourhood stars observed with IGRINS using the same line list and analysis technique; we also compare our sample to APOGEE DR17 abundances for inner bulge stars. We investigate bulge membership using 
   spectro-photometric distances and orbital simulations. We  construct a chemical-evolution model that fits our metallicity distribution function (MDF) and our $\alpha$-element trends.}
   {Among our 72 stars, we find four that are not bulge members. \sife\ and \mgfe\ versus \feh\ trends show a typical thick disc $\alpha$-element behaviour, except that we do not see any plateau at supersolar metallicities as seen in other works. The NLTE analysis lowers [Mg/Fe] typically by $\sim$0.1 dex, resulting in a noticeably lower trend of \mgfe\ versus \feh. The derived \cafe\ versus \feh\ trend has
   a larger scatter than those for Si and Mg, but is in excellent agreement with local thin and thick disc trends.
   With our updated analysis, we constructed one of the most detailed studies of the $\alpha$ abundance trends of cool M giants in the inner Galactic bulge. We modelled these abundances by adopting a two-infall chemical-evolution model with two distinct gas-infall episodes with timescales of 0.4\,Gyr and 2\,Gyr, respectively.}
   {  Based on  a very meticulous spectral analysis, we have constructed detailed and precise chemical abundances of Mg, Si, and Ca for cool M giants. The present study can be used as a  benchmark for future spectroscopic surveys. }


   \maketitle

\section{Introduction}

In  order to understand the formation and evolution of galaxies, we must begin by studying our own. 
To this purpose, the kinematics, dynamics, and chemical composition of stars are analysed mainly using orbit simulations and spectroscopy in order to set constraints on their formation processes. The different chemical elements observed in stellar spectra are created and distributed into the Galaxy by a large variety of astronomical processes, from nucleosynthesis in stars to that in supernovae. Among these elements, the $\alpha$-elements are the most commonly used for probing the stellar populations in the Galaxy. They are spread into the interstellar medium by means of core-collapse supernovae, which are violent events due to the death of the short-living massive stars. On the contrary, iron is mainly disseminated via thermonuclear supernovae, which happen on a  longer timescale. This longer timescale means that the enrichment of the interstellar gas varies with time and thus the $\alpha$-abundances found in stars depend on the time of their birth. $\alpha$-elements are therefore useful for determining the spatial (e.g. in situ or accreted) and temporal origins of stars; see for example \cite{vintergatan1},\cite{vintergatan2} and \cite{vintergatan3}.

In addition to the already well analysed Galactic discs, the inner regions of the Milky Way, ---the bulge, the nuclear stellar disc, and the nuclear star cluster--- are of specific interest in the quest to attain a complete picture of Galactic history. However, dust extinction towards the inner regions has prevented studies of these regions. More studies are therefore needed to unveil the details of the various stellar populations occupying these structures. In the study of the Galactic bulge\footnote{Here defined purely geometrically as the inner 10 degrees of the Milky Way \citep{barbuy:18}.},  industrial spectroscopy in both the optical and IR has become commonplace during the last decade, with surveys such as Gaia-ESO \citep{gaia-ESO:22}, APOGEE \citep{APOGEEDR17}, and GALAH \citep{galah:21}. These investigations have been powerful, especially in producing large numbers of  
abundance and composition measurements.
However, nearly all industrial spectroscopy projects forego abundance analysis and spectrum synthesis for individual stars, instead developing a range of tools from label matching to machine learning, which exploit high-resolution spectra and speed up their analysis via an automated pipeline.
Interestingly, some of the abundance trends and major conclusions, while showing the overall declining trend of [$\alpha$/Fe] versus [Fe/H], also reveal significant differences in shape; for example, \cite{queiroz2020} show two dominant trends: an alpha-enhanced thick disc trend, and a low alpha population that predominates in the central kiloparsec.  

The original hypothesis that the bulge is alpha-enhanced relative to the thin and thick disc \citep{mcwilliam1994,lecureur2007,zoccali2008,fulbright2007,johnson2014} has begun to yield to abundance trends that appear increasingly similar to a thick disc that extends to higher metallicity, $\sim +0.5$ dex \citep{melendez:08,Ryde2010,Rojas2017,Zasowski2019}. A general picture has emerged according to which the dynamical bar becomes dominant at [Fe/H]$>-0.5$ dex \citep{soto2007,ness2013} and the metal-rich population is concentrated in the plane \citep{ness2013,johnson2020,johnson2022}. New data have therefore challenged the early paradigm that the bulge formed early, rapidly, and with high $\rm[\alpha/Fe]$  set by \cite{matteucci&brocato} and \cite{mcwilliam1994}. If the bulge/bar abundance trends are understood as resembling those of the thick disc but extended to +0.5 dex, this is an important paradigm shift that has implications for how we consider the formation history of the bar and its relationship to the disc.

There is accumulating evidence in support of the view that the bulge contains at least two different structures or populations: (i) a metal-rich population that is most likely associated with the bar and therefore with a secular formation origin from the early disc (see e.g. \citealt{Ness2012}, \citealt{Rojas2017,Rojas2020}, \citealt{Zoccali2017}) and (ii) a metal-poor population that dominates away from the plane \citep{johnson2020,johnson2022} and whose formation scenario is less clear; suggestions include association with an early merger event, metal-enriched outflows, or secular evolution of the thick disc.

These paradigm shifts regarding the bulge challenge us not only to increase the sample size of abundance measurements, but also to reconsider making the most careful possible abundance determinations on smaller samples, particularly in the IR, where the most immediate advances in high-resolution spectroscopy are anticipated.

High-resolution, detailed abundance measurements near the plane and Galactic centre remain relatively sparse because of very high extinction. So far, the great majority of observations have been made in regions of relatively low extinction; for example, Baade's window, where many optical and IR studies exist. APOGEE used this window to obtain a high-quality set of chemical abundances \citep{Schultheis2017}. This is important if one wants to compare chemical abundances obtained by different instrument setups (e.g. wavelength range, spectra resolution, S/N, etc.) as each of those can introduce significant bias. However, a similar work but closer to the Galactic plane is missing. \cite{nandakumar:18} provide the largest sample of stars with high-resolution spectroscopic measurements near the plane, with their work finding no asymmetry between the northern and southern bulge, and confirming the presence of a metal-poor peak near -0.5 dex. 

Even in the era of industrial-scale spectroscopy, the development of small, high-quality samples analysed with different approaches has value, and especially in regions of high extinction and crowding near the plane. In this work, we therefore make use of the data sets provided by \cite{ryde:15}, \cite{ryde:16} and \cite{nandakumar:18} and reanalyse them consistently with updated methods. We apply new distance determinations and models of orbits, and use NLTE spectrum synthesis to give the best possible modern composition measurements of this sample. We additionally take advantage of new high-resolution, high-S/N IGRINS spectra of thin and thick disc stars in the solar neighbourhood. We report new trends in Si, Mg, and Ca in the inner galaxy, and fit the [Fe/H] distribution with chemical-evolution models.


\section{Observations}

We analysed high-resolution, near-IR spectra of stars in both the inner Galactic bulge and a local comparison sample. Observations were performed with the CRIRES and ISAAC spectrographs on the {Very Large Telescope} (VLT), the SOFI spectrographs on the {New Technology Telescope} (NTT), and the IGRINS spectrograph on the 4.3 m Lowell Discovery Telescope and the 2.7m McDonald telescope.

\subsection{Inner bulge sample}

The high-resolution, near-IR spectra of the 72 inner bulge giants analysed here for stellar abundances were observed in 2012-2013 with CRIRES \citep{crires:04,crires:05,crires:06} mounted on the VLT, and are described in detail in \citet{nandakumar:18} (Paper I). A relatively short wavelength coverage was observed in the K band, namely $20\,818-21\,444$\,\AA, but at a spectral resolving power of $R\sim50\,000$. 

The effective temperatures of the stars needed for the analysis were determined in Paper I from low-resolution, K-band spectra ($R\sim1000-2000$). These spectra were observed with the ISAAC  \citep[][]{isaac} on the VLT and the SOFI spectrographs \citep{sofi} on ESO's NTT telescope on La Silla. 

Stars in six fields along the minor axis, both south and north of the Galactic centre, were observed at both high and low spectral resolution; see Table \ref{table:stellarparam_N} and \ref{table:stellarparam_S}. Among the 72 M giants, 44 belong to northern fields, 19 to southern fields, and 9 to a centre field.   The stars in the latter field lie $2.5-5.5$ arcmin north of the very centre of the Milky Way ---which corresponds to a projected galactocentric distance of $5-10$ pc--- and probably belong to the nuclear disc and not the nuclear star cluster \citep[cf.][]{ryde:16:metalpoor,rich:17}.

\begin{figure}[ht!]
  \centering
  \includegraphics[width=\hsize]{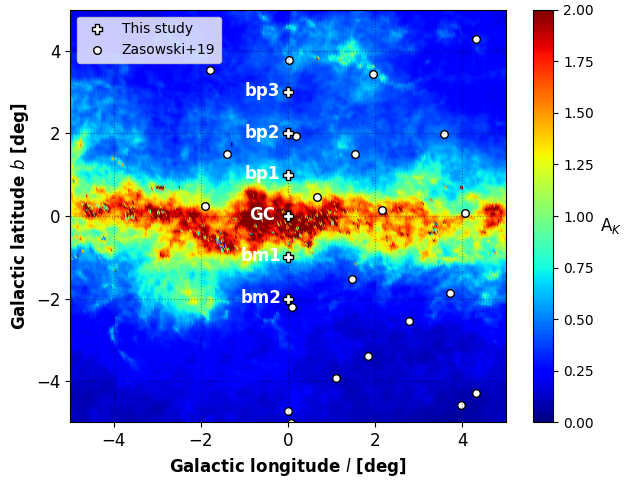}
      \caption{Location of the inner Galactic bulge fields studied in this work and those from \cite{Zasowski2019} (only those visible in this frame) in Galactic coordinates superimposed on the interstellar extinction map of \cite{Gonzalez2012}.}
         \label{fig:lb}
\end{figure}

\subsection{Solar neighbourhood sample}
\label{sec:obs_SN}

In order to minimise possible systematic uncertainties in the analysis of the abundance trends for the inner Galactic bulge, we observed similar types of stars in the Galactic discs as locally as possible. With determined abundance trends from these stars, we can search and study possible differences in a differential way.  This comparison sample of 37 solar neighbourhood K giants was observed with the Immersion GRating INfrared Spectrograph  \citep[IGRINS;][]{Yuk:2010,Wang:2010,Gully:2012,Moon:2012,Park:2014,Jeong:2014} in 2016 on the 4.3 m Discovery Channel Telescope (DCT; now called the Lowell Discovery Telescope) at Lowell Observatory \citep{Mace:2018}, and on the 2.7 m Harlan J. Smith Telescope at McDonald Observatory \citep{Mace:2016}. The near-IR spectra provided by the IGRINS span the full H and K bands (1.45 - 2.5 $\mu$m) with a spectral resolving power of R $\sim$ 45000. Details of observations and data reduction are explained in detail in \cite{Ryde:2020}, \cite{montelius:22} and \cite{Nandakumar:2022}.


\section{Analysis}
\label{sec:analysis}

In this study, we determined the abundance trends versus metallicity of the $\alpha$-elements Mg, Si, and Ca for the inner bulge sample, as well as the local comparison sample with the same method. We made an effort to clarify the membership of the stars in our inner bulge sample by determining accurate distances and performing a dynamic analysis of the stars. In this way, we were able to disregard any star in the sample that is obviously not a bulge member. 
Furthermore, we adopted an updated K-band line list with determined astrophysical line strengths and improved van der Waals as well as Stark broadening parameters.
We also applied and discuss non-LTE corrections for all spectral lines.  

\subsection{Spectroscopic analysis}

\subsubsection{Stellar parameters}
\label{sec:stellarparam}

Proper determination of the stellar parameters (i.e. \teff, \logg, \feh, and \vmic) of the stars is fundamental for an accurate and precise abundance determination. The effective temperatures used in this paper were determined in Paper I based on the relation between the effective temperature and the $^{12}$CO-band-head strength in low-resolution spectra at $2.3\,\mu$m. The method is described in \citet{teff} and results in typical uncertainties of $\pm150$K.
Furthermore, first estimations of the metallicities are determined using \logg\ and \vmic\ from Paper I. Given these metallicities, new surface gravities and microturbulences were then determined with the iterative method from \citet{rich:17} and the empirical relation between microturbulence and surface gravity from \citet{2013ApJ...765...16S}, respectively. Typical uncertainties in surface gravity are of $\pm0.3$\,dex after a few iterations.

The line shapes and widths are important in our analysis because we compare synthesised lines with observed ones. Therefore, we carefully analyse the widths of the lines and estimate a general macroturbulence, \vmac, for each star using manually selected Gaussian-shaped spectral lines from Mg, Al, and Si. The updated broadening parameters for these lines will be important for the stronger lines.

The new stellar parameters are given in Tables \ref{table:stellarparam_N} and \ref{table:stellarparam_S}. For our star sample, \teff\ is in the range $3400$ - $4400$ K and \logg\ is within $0.4$ - $2.0$ dex.
Uncertainties in the derived abundances\footnote{The solar value used in this study is from \cite{grevesse2007}} are due to the fitting procedure and to the uncertainties on the stellar parameters (as shown in \cite{rich:17,ryde:16}) and are already well discussed in Section 3.4 of Paper I. It has been shown that the fitting procedure is the main contributor to the abundance error. Therefore, as in Paper I, we adopt a main abundance uncertainty of $\pm0.15$dex.

\begin{table}
\caption{Stellar parameters of the observed stars in the northern fields of the inner Galactic bulge. The names of stars belonging to our Golden Sample (see Section \ref{sec:golden sample}) are marked in bold face.}\label{table:stellarparam_N}
\begin{tabular}{|l|c|c|c|c|c|}
\hline
Star & \teff & \logg & \feh\ &  $\xi_\mathrm{macro}$ & $\xi_\mathrm{micro}$\\
         & [K] & (cgs) &  & [km\,s$^{-1}$] & [km\,s$^{-1}$]\\  
\hline
\multicolumn{6}{|l|}{Northern field at $(l,b) = (0^{\circ},+3^{\circ})$}\\
\hline
 \textbf{bp3-01} &        3780 &       0.75 &      -0.67 &     4.9 &    2.2 \\
 \textbf{bp3-02} &        4111 &       1.83 &       0.08 &     4.2 &    1.8 \\
 \textbf{bp3-04} &        3623 &       0.67 &      -0.34 &     6.0 &    2.3 \\
 \textbf{bp3-05} &        3879 &       1.67 &       0.50 &     5.9 &    1.8 \\
 bp3-06          &        3755 &       0.63 &      -0.78 &     3.7 &    2.3 \\
 \textbf{bp3-07} &        3962 &       1.77 &       0.40 &     5.5 &    1.8 \\
 \textbf{bp3-08} &        3637 &       1.04 &       0.20 &     6.0 &    2.1 \\
 \textbf{bp3-10} &        4052 &       1.74 &       0.12 &     4.4 &    1.8 \\
 \textbf{bp3-11} &        3542 &       0.69 &      -0.06 &     6.4 &    2.3 \\
 \textbf{bp3-12} &        3966 &       1.36 &      -0.21 &     6.4 &    1.9 \\
 \textbf{bp3-13} &        3625 &       0.84 &      -0.05 &     7.2 &    2.2 \\
 \textbf{bp3-14} &        3569 &       0.63 &      -0.26 &     6.4 &    2.3 \\
 \textbf{bp3-15} &        3610 &       0.53 &      -0.55 &     5.9 &    2.4 \\
 \textbf{bp3-16} &        3654 &       0.84 &      -0.13 &     7.0 &    2.2 \\
 \textbf{bp3-17} &        3946 &       1.35 &      -0.16 &     5.3 &    1.9 \\
\hline
 \multicolumn{6}{|l|}{Northern field at $(l,b) = (0^{\circ},+2^{\circ})$}\\
\hline
 \textbf{bp2-01} &        4054 &       1.93 &       0.38 &     5.6 &    1.8 \\
 bp2-02          &        3838 &       1.31 &       0.08 &     5.6 &    1.9 \\
 \textbf{bp2-03} &        3889 &       1.45 &       0.14 &     6.6 &    1.9 \\
 \textbf{bp2-04} &        4134 &       1.60 &      -0.32 &     4.9 &    1.9 \\
 \textbf{bp2-05} &        4079 &       1.36 &      -0.55 &     5.0 &    1.9 \\
 bp2-06          &        3962 &       0.47 &      -1.80 &     5.6 &    2.4 \\
 \textbf{bp2-07} &        4320 &       1.77 &      -0.59 &     4.0 &    1.8 \\
 \textbf{bp2-08} &        3808 &       1.29 &       0.13 &     6.0 &    2.0 \\
 \textbf{bp2-09} &        3813 &       0.79 &      -0.69 &     4.0 &    2.2 \\
 \textbf{bp2-10} &        3774 &       1.29 &       0.20 &     5.2 &    2.0 \\
 \textbf{bp2-11} &        3645 &       1.10 &       0.26 &     6.3 &    2.0 \\
 \textbf{bp2-12} &        4032 &       1.17 &      -0.71 &     7.6 &    2.0 \\
 \textbf{bp2-13} &        3899 &       1.10 &      -0.45 &     4.2 &    2.0 \\
 \textbf{bp2-14} &        3559 &       0.85 &       0.11 &     5.5 &    2.2 \\
 \textbf{bp2-15} &        4237 &       1.51 &      -0.74 &     5.7 &    1.9 \\
\hline
 \multicolumn{6}{|l|}{Northern field at $(l,b) = (0^{\circ},+1^{\circ})$}\\
\hline
 bp1-01          &        3918 &       1.69 &       0.40 &     5.6 &    1.8 \\
 \textbf{bp1-02} &        4311 &       1.47 &      -0.99 &     5.0 &    1.9 \\
 \textbf{bp1-03} &        3948 &       1.79 &       0.49 &     7.5 &    1.8 \\
 \textbf{bp1-04} &        4060 &       1.67 &      -0.01 &     5.9 &    1.8 \\
 \textbf{bp1-05} &        3731 &       1.34 &       0.37 &     5.2 &    1.9 \\
 \textbf{bp1-06} &        3863 &       1.18 &      -0.18 &     6.3 &    2.0 \\
 \textbf{bp1-07} &        4209 &       1.47 &      -0.73 &     3.9 &    1.9 \\
 \textbf{bp1-08} &        4242 &       1.65 &      -0.56 &     5.3 &    1.9 \\
 \textbf{bp1-09} &        3494 &       1.00 &       0.49 &     4.9 &    2.1 \\
 \textbf{bp1-10} &        3879 &       1.40 &       0.10 &     5.6 &    1.9 \\
 \textbf{bp1-11} &        4352 &       1.94 &      -0.42 &     5.4 &    1.8 \\
 \textbf{bp1-12} &        4175 &       1.91 &       0.01 &     5.2 &    1.8 \\
 \textbf{bp1-13} &        4151 &       1.45 &      -0.60 &     5.3 &    1.9 \\
 \textbf{bp1-14} &        4070 &       1.63 &      -0.10 &     4.2 &    1.9 \\
\hline
\end{tabular}
\end{table}

\begin{table}
\caption{Stellar parameters of the observed stars in the southern fields of the inner Galactic bulge. The names of stars belonging to our Golden Sample (see Section \ref{sec:golden sample}) are marked in bold face.}\label{table:stellarparam_S}
\begin{tabular}{|l|c|c|c|c|c|}
\hline
\multicolumn{1}{|l|}{Star} & \teff & \logg & \feh &  $\xi_\mathrm{macro}$ & $\xi_\mathrm{micro}$\\
         & [K] & (cgs) &  & [km\,s$^{-1}$] & [km\,s$^{-1}$] \\  
\hline
\multicolumn{6}{|l|}{Galactic Centre field at $(l,b) = (0^{\circ},0^{\circ})$}\\
\hline
 GC1             &        3668 &       1.12 &       0.23 &     5.6 &    2.0 \\
 GC20            &        3683 &       1.25 &       0.36 &     8.3 &    2.0 \\
 \textbf{GC22}   &        3618 &       1.13 &       0.36 &     4.9 &    2.0 \\
 GC25            &        3340 &       0.67 &       0.37 &     6.6 &    2.3 \\
 GC27            &        3404 &       0.68 &       0.26 &     5.5 &    2.3 \\
 GC28            &        3773 &       1.34 &       0.28 &     6.1 &    1.9 \\
 \textbf{GC29}   &        3420 &       0.86 &       0.46 &     6.4 &    2.2 \\
 GC37            &        3754 &       1.42 &       0.44 &     7.0 &    1.9 \\
 \textbf{GC44}   &        3465 &       0.96 &       0.49 &     4.9 &    2.1 \\
\hline
 \multicolumn{6}{|l|}{Southern field at $(l,b) = (0^{\circ},-1^{\circ})$}\\
\hline
 bm1-06          &        3814 &       1.54 &       0.46 &     6.1 &    1.9 \\
 \textbf{bm1-07} &        3873 &       1.44 &       0.17 &     5.1 &    1.9 \\
 \textbf{bm1-08} &        3650 &       1.21 &       0.39 &     6.0 &    2.0 \\
 \textbf{bm1-10} &        3787 &       1.10 &      -0.09 &     4.9 &    2.0 \\
 \textbf{bm1-11} &        3812 &       1.48 &       0.38 &     6.8 &    1.9 \\
 bm1-13          &        3721 &       0.44 &      -0.93 &     4.9 &    2.4 \\
 bm1-17          &        3775 &       0.56 &      -0.91 &     5.6 &    2.4 \\
 bm1-18          &        3780 &       1.48 &       0.46 &     5.6 &    1.9 \\
 bm1-19          &        3958 &       1.76 &       0.40 &     5.3 &    1.8 \\
\hline
 \multicolumn{6}{|l|}{Southern field at $(l,b) = (0^{\circ},-2^{\circ})$}\\
\hline
  \textbf{bm2-01} &        3946 &       1.51 &       0.08 &     3.5 &    1.9 \\
 \textbf{bm2-02} &        4013 &       1.44 &      -0.22 &     6.7 &    1.9 \\
 \textbf{bm2-03} &        3668 &       1.24 &       0.39 &     6.0 &    2.0 \\
 \textbf{bm2-05} &        3450 &       0.86 &       0.39 &     6.1 &    2.2 \\
 \textbf{bm2-06} &        4208 &       1.29 &      -0.97 &     5.1 &    2.0 \\
 \textbf{bm2-11} &        4005 &       1.16 &      -0.65 &     6.0 &    2.0 \\
 \textbf{bm2-12} &        4003 &       1.73 &       0.23 &     6.5 &    1.8 \\
 \textbf{bm2-13} &        3727 &       1.05 &       0.00 &     5.0 &    2.1 \\
 \textbf{bm2-15} &        3665 &       1.31 &       0.52 &     6.3 &    1.9 \\
 \textbf{bm2-16} &        3886 &       1.58 &       0.32 &     4.7 &    1.9 \\
\hline
\end{tabular}
\end{table}

\subsubsection{Line list}
\label{sec:linelist}

In order to determine reliable abundances for different elements from stellar spectra, we need information about the atomic physics data for the spectral lines corresponding to each element. We need the central wavelength of the line, upper and lower excitation energies of transition levels, line strengths (log$gf$), broadening parameters, and so on. The line list has all this information, which is necessary in order to synthesise spectral lines. VALD \citep{vald:2015} assimilates this information from a large collection of sources, which we use as our base line list in this work. 

However, we update the line list primarily with log$gf$ values for our lines of interest (without reliable experimental log$gf$ values), which we determined astrophysically using the high-resolution IR solar flux spectrum of \cite{Wallace:2003}. We do this by finding the value of log$gf$ for which the respective line in the solar-flux spectrum is best fitted by the synthetic spectrum.  

The collisional broadening due to neutral hydrogen, and in some cases charged particles, is important for strong lines with damping wings.  For collisional broadening due to hydrogen, where possible, we have used parameters from the ABO theory \citep{anstee_investigation_1991,Anstee1995,Barklem1997a,Barklem1998b} taken from the spectral synthesis code BSYN based on routines from MARCS \citep{marcs:08}.
For the \ion{Mg}{i} lines at 21059.75 \AA\, and the multiplet at 21060.71 - 21061.095 \AA,\, the parameters used in Paper I for collisional broadening due to hydrogen were $\sigma(v=10^4\mbox{m/s}) = 2927 \, a_0^2$, with velocity parameter $\alpha=1.313$\footnote{Here $\sigma$ is the broadening cross section and $\alpha$ describes the velocity dependence assuming a power law $\sigma \propto v^{-\alpha}$.}. These values were calculated using the impulse approximation \citep{Kaulakys1985,Kaulakys1991,HoangBinh1995,osorio_mg_2015} based on Coulomb wave functions in momentum space \citep{HoangBinh1997, barklem_mswavef_2015}, but neglecting elastic contributions because of the problem of estimating the interference between the lower and upper level contributions. In this work, rather than neglecting the elastic contribution, we estimated the maximum and minimum possible contributions, that is, the completely constructive and destructive interference cases, respectively.  
The mean of these two extreme elastic contributions was then added to the inelastic contribution, and we obtained $\sigma(v=10^4\mbox{m/s}) \approx 4440 \, a_0^2$ and $\alpha=1.10$.  The error in the cross section due to the uncertainty in the elastic contribution is roughly 15\%, with the overall uncertainty from this method probably being at least of order 30\%.

Further, for the 21059~\AA\ line, data for Stark broadening were extracted from the Stark-B database \citep{sahal-brechot_starkb_2015}, the data originating from \cite{Dimitrijevic1996}.  Combining the data for broadening due to electrons ($e$) and protons ($p$) with the assumption $N_p=N_e$, and converting to the format conventionally used in many spectrum-synthesis codes for quadratic Stark broadening, we find that $\log \Gamma/N_e = -2.56$ at $T=10000$~K, where $\Gamma$ is the full-width at half maximum in $\mbox{rad}/\mbox{s}$ and $N_e$ is in cm$^{-3}$.  We find no data for the 21060.71 - 21061.095 \AA\ multiplet.  As this multiplet is analogous to the 21059~\AA\ transition but in the triplet term system, we assume the Stark broadening for this multiplet to be the same as the line in the singlet system.  The combination of the increased collisional broadening due to hydrogen and the strong Stark broadening significantly improved the agreement with the observed spectrum in the Sun.

In addition to the Sun spectrum, we used the high-resolution (R $\sim$ 100,000) IR spectrum of the well-studied Arcturus from the Arcturus atlas \citep{Hinkle:1995} to validate the updated line strengths of Fe, Mg, Si, and Ca lines. Arcturus is a well-studied reference star with reliable stellar parameters determined using various methods and is a cool giant similar to the stars investigated in this work. We determined mean values of -0.5, 0.22, 0.27, and 0.29 dex for [Fe/H] (excluding four Fe lines that were noisy or telluric affected in the Arcturus spectrum), [Mg/Fe], [Si/Fe], and [Ca/Fe], respectively, with less than 0.05 dex line-by-line dispersion for each element assuming the stellar parameters from \cite{ramirez:2011}. In addition, we used the same line list to determine the elemental abundances of the 37 solar neighbourhood giants using the same set of lines (see Section~\ref{sec:analysis_SN} and upper panels of Figure~\ref{fig:sn&zasowski}).
  
  The central wavelengths, updated log$gf$ values, and broadening parameters for the spectral lines used in this work are listed in Table~\ref{table:lines}. For the  molecular lines, we used the line data for CO from \citet{li:2015}, for CN from \citet{sneden:2014}, and for OH from \citet{brooke:2016}.

\begin{table}
\caption{Spectral line data for the lines used in the present study. }\label{table:lines}
\begin{tabular}{|l| c| c| c| c|}
\hline
 Element & Wavelength  & log (gf) &  Broadening &  $\alpha$\\
 & ($\AA$)  & &   by H ${^a}$  &  \\
\hline
\multirow{9}{*}{Si} & 20804.225 $^{1 *}$   & -1.026 $^{1 *}$   & 861 $^{2}$   &  0.292  \\
 &  20890.415 $^{1 *}$   & -1.613 $^{1 *}$   & 859 $^{2}$   &  0.292  \\
 &  20908.625 $^{1}$   &  -1.453 $^{1 *}$  & -6.990 $^{1}$   &    \\
 &  20917.151 $^{1}$   &  0.288 $^{1 *}$  & 1485 $^{2}$  & 0.324   \\
 &  20926.149 $^{1}$   &  -1.076 $^{1 *}$  &  1484 $^{2}$ & 0.324     \\
 &  21056.379 $^{1 *}$   &  -0.505 $^{1 *}$  &  -7.032 $^{1 *}$  &    \\
 &  21139.759 $^{1}$   &  -0.501 $^{1 *}$  & -6.880 $^{1 }$   &    \\
 &  21143.260 $^{1 *}$   &  -0.619 $^{1 *}$  & -6.880 $^{1 }$   &    \\
 &  21204.492 $^{1 *}$   &  -0.388 $^{1 *}$  & -6.880 $^{1 }$   &    \\
\hline
\multirow{2}{*}{Mg} & 21059.757 $^{4 *}$   &  -0.384 $^{5 *}$  &  4440 $^{3}$  &  1.10  \\
&  21060.710 $^{4 *}$   & -0.530 $^{5 }$   & 4440 $^{3}$   &   1.10  \\
&  21060.896 $^{4 *}$   & -1.587 $^{5 }$   & 4440 $^{3}$   &    1.10 \\
&  21060.896 $^{4 *}$   & -0.407 $^{5 }$   & 4440 $^{3}$   &   1.10  \\
&  21061.095 $^{4 *}$   & -3.383 $^{5 }$   & 4440 $^{3}$   &   1.10  \\
&  21061.095 $^{4 *}$   & -1.583 $^{5 }$   & 4440 $^{3}$   &    1.10 \\
&  21061.095 $^{4 *}$   & -0.298 $^{5 }$   & 4440 $^{3}$   &   1.10  \\
\hline
\multirow{3}{*}{Ca} 
 &  20962.570 $^{1 *}$   &  -0.784 $^{1 *}$  &  -7.230 $^{1 }$  &    \\ 
 &  20972.529 $^{1 }$   &  -1.002 $^{1 *}$  &  -7.230 $^{1 }$  &    \\ 
 &  20973.378 $^{1}$   &  -1.436 $^{1 *}$  &  -7.230 $^{1 }$  &    \\ 
\hline
\multirow{12}{*}{Fe} & 20798.893 $^{6}$ & -3.525 $^{6}$ & -7.280 $^{6 }$ &    \\
 & 20799.685 $^{6}$ & -0.613 $^{6*}$ & -7.021 $^{6*}$ &    \\
 & 20805.113 $^{7 *}$ & -0.065 $^{7 *}$ & -7.076 $^{7 *}$ &    \\
 & 20840.835 $^{6 *}$ & 0.104 $^{6 *}$ & -7.134 $^{6 *}$ &    \\
 & 20882.233 $^{6 }$ & -0.935 $^{6 *}$ & -7.330 $^{6 }$ &    \\
 & 20948.086 $^{6}$ & -0.796 $^{6 *}$ & -6.861 $^{6 *}$ &    \\
 & 20991.083 $^{6 *}$ & -3.058 $^{6 *}$ & -7.730 $^{6 }$ &    \\
 & 21036.355 $^{6}$ & -0.817 $^{6 *}$ & 2656 $^{2}$ &  0.330  \\
 & 21095.401 $^{6}$ & -0.658 $^{6 *}$ & -7.000 $^{6 *}$ &    \\
 & 21105.217 $^{6}$ & -0.749 $^{6 *}$ & -7.320 $^{6 }$ &    \\
 & 21124.505 $^{6 *}$ & -1.647 $^{6 *}$ & 975 $^{2}$ &  0.302  \\
 & 21162.095 $^{6 *}$ & -0.605 $^{6 *}$ & 1364 $^{2}$ &  0.328  \\
 
\hline
\end{tabular}
\tablefoot{a: Collisional broadening by neutral hydrogen gives either the broadening cross section $\sigma(v=10^4\mbox{m/s})$ in atomic units ($a_0^2$), with velocity parameter $\alpha$ (see text for more details), or if the value is negative and no $\alpha$ given, then the line width is given in the standard form  $\log \Gamma/N_H$ at $T=10000$~K, where $\Gamma$ is the full-width at half maximum in $\mbox{rad}/\mbox{s}$ and $N_H$ is in cm$^{-3}$. 1: \cite{K07}, 2: BSYN based on routines from MARCS code \cite{marcs:08}, 3: see text, 4: \cite{Brault:1983}, 5: \cite{civis:2013}, 6: \cite{K14}, 7: manual entry by NIST lookup, *: astrophysical estimate }
\end{table}

\begin{figure*}[ht!]
   \centering
   \includegraphics[width=\hsize]{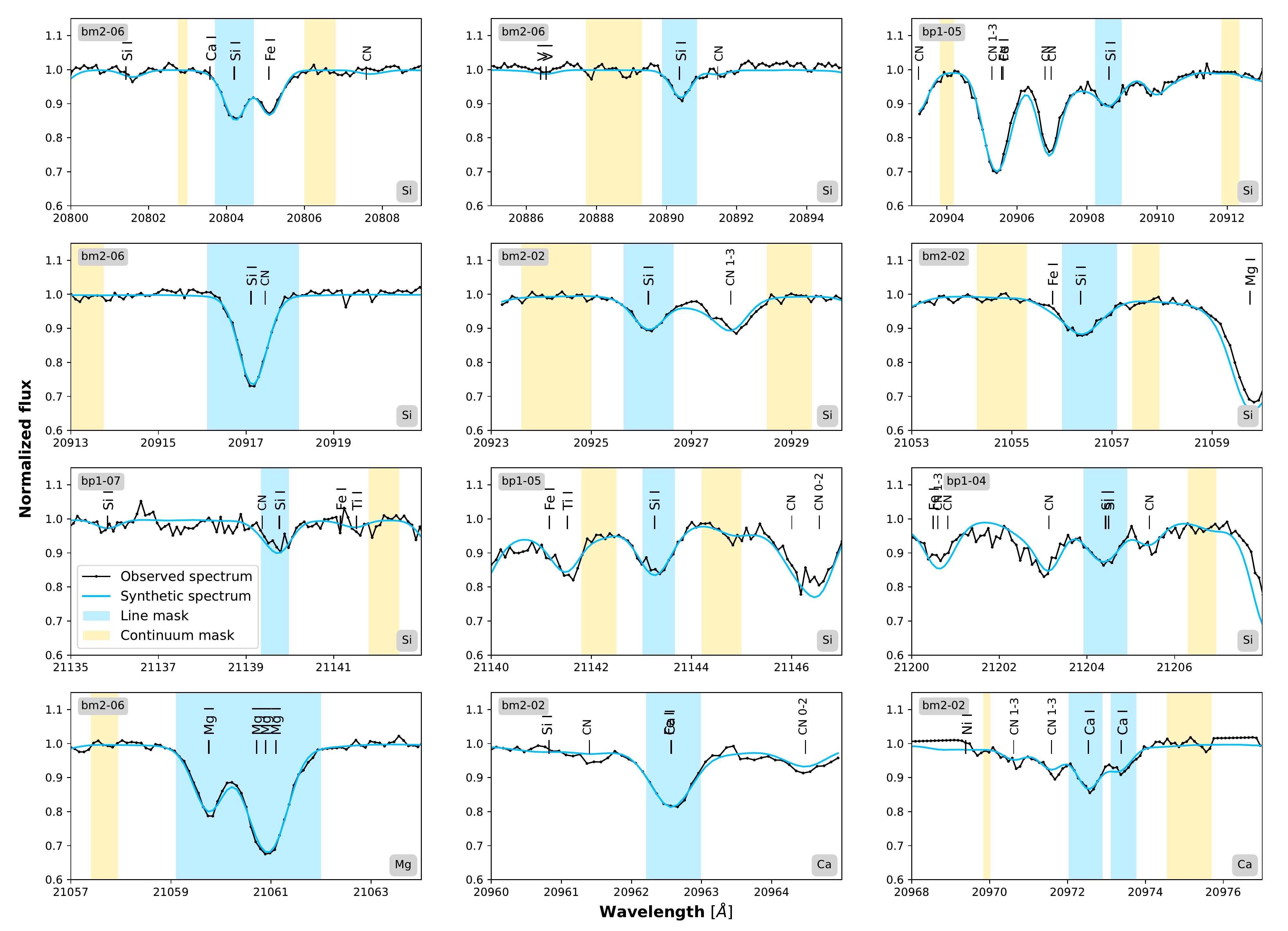}
      \caption{Typical spectral fits for Si (the nine upper plots), Mg (lower left), and Ca lines (lower middle and right) used in the analysis.  The stellar name is given in the upper left corner of every subplot. The black observed data are fitted with the blue synthetic spectrum. The lines of interest are marked with the blue line masks. The yellow masks show the regions in the spectrum that are used to normalise the local continuum around the lines. In some of the plots, these continuum regions are outside of the figure. The linear local continuum is always fitted with at least three continuum windows in every segment close to the line of interest.}
         \label{fig:spectra all si mg ca lines}
\end{figure*}

\begin{figure*}[ht!]
   \centering
   \includegraphics[width=\textwidth]{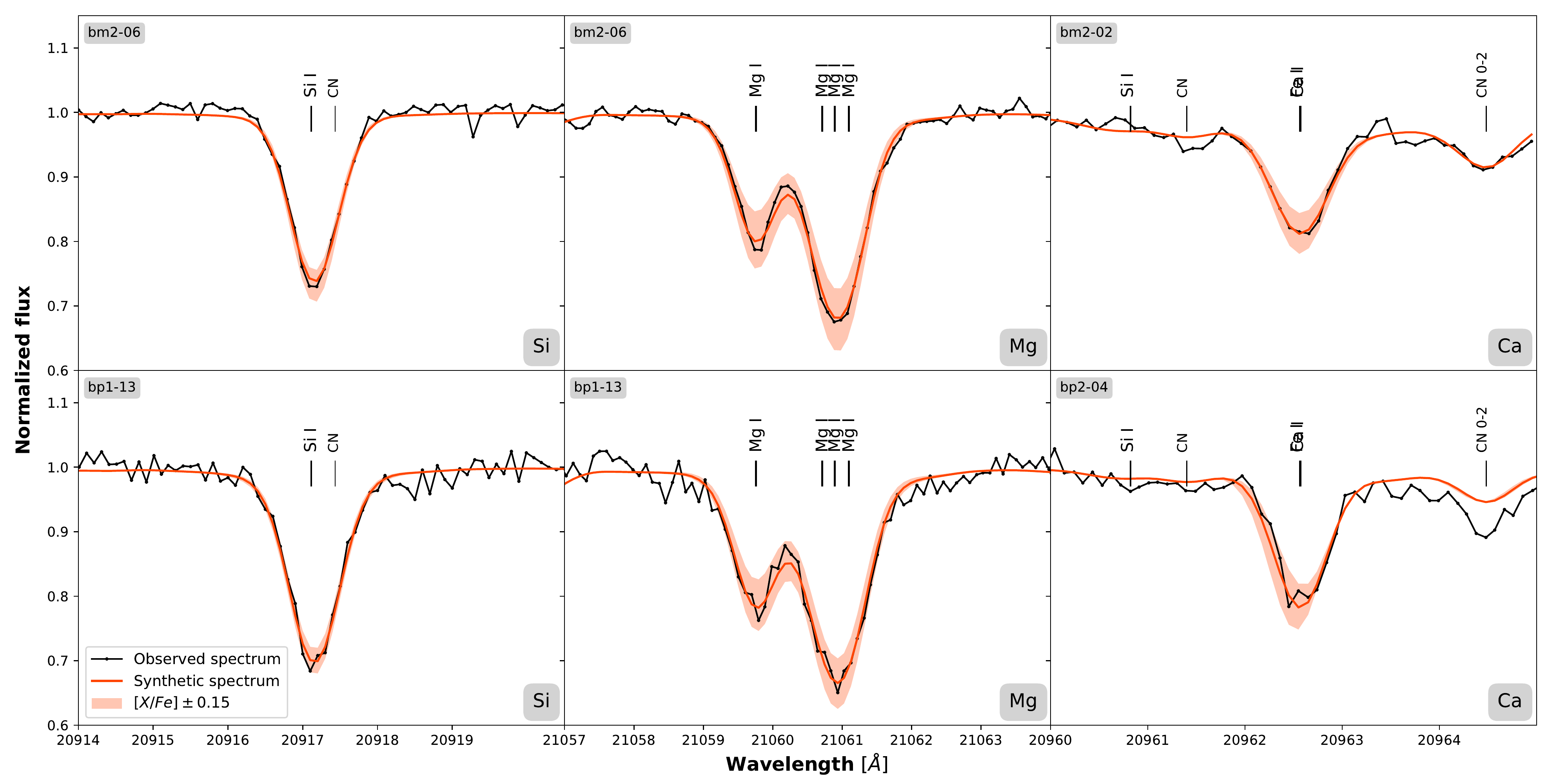}
      \caption{Typical spectral fits for Si, Mg, and Ca lines with a variation band of $\pm0.15$ dex in abundance. The upper panel shows fitted lines for stars with a good S/N whereas the lower panel shows those for stars with a poor S/N.}
         \label{fig:spectra fit example}
\end{figure*}

\subsubsection{Spectral synthesis}
\label{sec:spectral analysis}

We used Spectroscopy Made Easy (SME version 583, \citet{sme1,sme2}) for our spectral analysis. For a given set of stellar parameters, in order to generate synthetic spectra, 
SME interpolates a model atmosphere from a grid of spherically symmetric MARCS models computed using LTE.
To account for non-LTE effects, SME applies departure coefficients by interpolating in grids. In this study, we used this type of non-LTE grid for which the departure coefficients were computed using the MPI-parallelised non-LTE radiative transfer code \texttt{Balder}. This code is based on the code \texttt{Multi3D} \citep{multi3D}, which calculates the level populations for model atoms  based on model atmospheres, forcing a statistical equilibrium for all levels in the ambient radiation field. For a further description of the general approach to the calculations, see \cite{NLTE}. Specifically, in our analysis we used the grids for Si, Mg, and Ca from \cite{NLTE} and for Fe from \cite{lind17} and \cite{amarsi16} (with subsequent updates Amarsi priv. comm.).  

Moreover, SME needs the user to select spectral intervals within which the studied spectral lines lie and a line list containing the atomic and molecular data for these segments (see more about our line list in Section \ref{sec:linelist}). Regions free from spectral lines and other features are marked manually as continuum masks allowing a local linear continuum normalisation within the interval. In the same way, the line of interest has to be marked with a line mask. Therefore, for measuring the abundance or other parameters, SME generates synthetic spectra and compares them with the observed spectrum using a $\chi^2$-minimisation in the region within the line masks.

As explained in Section \ref{sec:stellarparam}, for the abundance analysis, firstly we determined the metallicity \feh\ using the iterative method from \citet{rich:17}. For each chemical element studied, namely iron, silicon, magnesium, and calcium, we chose a set of good-quality spectral lines (the ones used in our study are shown in Figure \ref{fig:spectra all si mg ca lines} and listed in Table \ref{table:lines}). Among these, we meticulously selected the best available lines   for each star. To this end, each line and spectral interval were examined by eye so as to evaluate S/N and to detect disturbing artefacts or cosmic hits. In addition, line blending and fitting quality were appraised. In order to determine the final abundance value, the mean of the abundances derived from the selected lines is computed. This process permits us to obtain a precise estimate of the chemical abundances of our stars.

Given that metallicity is essential in the determination of stellar models and then in the estimation of the chemical fingerprints of stars, the iron abundances were directly determined with non-LTE corrections, contrary to the three other chemical elements for which LTE and non-LTE cases were explored and compared.

In the case of magnesium, we used the line at $21059.757\,\AA$ and the multiplet lines at $21060.710 - 21060.896\,\AA$. As shown in Fig. \ref{fig:spectra all si mg ca lines}, these lines are wide and very well defined and are therefore good candidates for the abundance determination as we have precise broadening values. Moreover, due to the proximity of these spectral features, we fitted all these lines simultaneously using one line mask.

Compared to the other elements of this study, there are only a few Ca lines available and these all lie on the edge of one of the detector chips. 
We  were therefore only able to obtain Ca abundance estimates for 42 stars of our sample.
Because of their proximity to CN lines, and in order to fit our selected Ca lines correctly, we fitted the nitrogen abundance via these molecular lines. Our aim is not to have an accurate estimation of nitrogen (which is a difficult chemical element to study because of its complicated molecular equilibrium with carbon and oxygen) but only to correctly fit the CN lines. 
Two additional Ca lines, located at $20956.394 \AA$ and $20962.570 \AA$, are heavily blended with CN lines, and are therefore insensitive to the Ca abundance. As a consequence, we discarded them from the study.

We ensured that all the spectral lines used were not blended with any other line of our line list. Even if no suspicious line has been retained, it should be noted that there is always a risk that unknown blends or lines with unknown line strengths are not accounted for, especially for metal-rich and cool stars. 

We chose a final sample, which we refer to as the Golden Sample (see Section \ref{sec:golden sample}), based on a sufficiently high S/N (> 40) and low line-by-line discrepancy ($\sigma_{lines}$ < 0.3 dex).


\subsubsection{Solar neighbourhood}
\label{sec:analysis_SN}

The 37 solar neighbourhood stars used for comparison in this work are a subset of the Giants In the Local Disk (GILD) stellar catalogue (J\"onsson et al. in prep.; which builds upon and improves the analysis described in \citet{jonsson:17}). We estimated the fundamental stellar parameters, effective temperature (\teff), surface gravity (\logg), metallicity (\feh), and  microturbulence ($\xi_\mathrm{micro}$) from carefully analysed optical spectra. We benchmarked these parameters  against independently determined effective temperatures, \teff, from angular diameter measurements and surface gravities, \logg, from asteroseismological measurements \citep[see][for more details]{jonsson:17} with uncertainties of $\pm$50K for \teff, $\pm$0.15 dex for \logg, $\pm$0.05 dex for \feh, and $\pm$0.1 km/s for $\xi_\mathrm{micro}$. Stellar parameters of these 37 stars are in the range of 4000 - 5100 K in \teff, 1 - 3.5 dex in \logg\, and -1 - 0.3 dex in \feh. Therefore, even though there is an overlap in \teff, the solar neighbourhood sample is strictly above 4000 K. There may therefore exist unknown systematic uncertainties in \teff\, which should be kept in mind while comparing the two samples. This is an improvement that could be made in the future when such observations exist. 

With accurate stellar parameters known, we carried out a detailed analysis of the same set of Si, Mg, and Ca lines used for the bulge stars in order to determine their respective abundances. In addition, we use the same set of stellar atmosphere grids, the same version of the SME, the same non-LTE corrections, and the same line list as we use for the bulge stars.  
We therefore carried out a consistent spectroscopic analysis of the bulge and solar neighbourhood giant stars to determine their elemental abundances. 
Needless to say, one obvious difference between the IGRINS spectra of the bright nearby solar neighbourhood stars and the CRIRES spectra of the faint and distant bulge stars  is the better S/N, which naturally affects the precision of the  abundance estimates.

\subsection{Dynamical analysis}
\label{sec:dynamics}

 While our sample size is relatively modest, we undertook a basic dynamical analysis in order to ensure membership of  the inner Galactic bulge. Therefore, astrometric parameters are necessary as initial conditions to study dynamics.
Among the 72 stars, 63 were cross-matched with Gaia EDR3 (\citealp{gaia2016,gaia2021}) in order to obtain their proper motions. To ensure reliable values, a proper motion error limit is set at $0.3$\,mas yr$^{-1}$. In addition, updated distances were determined (see Section \ref{sec:distances}) and radial velocities were measured using the strong and clear Mg lines.

We used the software package AGAMA \citep{agama} to compute the orbits of our stars up to 10 Gyr. As we are studying inner bulge stars, we used a combination of three gravitational potentials: a {disc}, a {halo,} and a {bulge/bar}. More precisely, we used the disk and halo components of the \texttt{MWPotential2014} \citep{galpy} and the bar potential from \cite{launhardtbar}. A typical (clock-wise) rotation pattern speed about $\Omega_{b}= 40$\,km s$^{-1}$ kpc$^{-1}$ was taken into account for the bar \citep{Portail17} which is at an angle of $\alpha = 25^{\circ}$ from the line of sight towards the Galactic centre. We use these orbital parameters to assign the bulge membership in addition to the calculated distances (see  Sect.~\ref{sec:membership})


\section{Results}
\label{sec:results}
We now discuss our new abundance measurements of $\alpha$ elements for giants in the inner Galactic bulge.
Moreover, we discuss the re-determined distances of the stars and therefore their bulge-membership.

\subsection{Abundances}

Figure \ref{fig:spectra fit example} shows typical fits of a Si line, Ca line, and the Mg lines. The sensitivity of the abundance determination is shown by varying the abundance by $\pm0.15$ dex (the choice of this uncertainty value is explained Section \ref{sec:stellarparam}) plotted as a band around the observed spectrum.

\subsubsection{LTE and non-LTE comparison}
\label{sec:LTEvsNLTE} 
In order to investigate the effects of non-LTE corrections and to be able to compare our results with those from Paper I, in which only the LTE case was explored, we determined Si, Mg, and Ca abundances calculated both with LTE and non-LTE level populations, affecting both the source function and the line opacity. Therefore, LTE and non-LTE abundance trends for \sife, \mgfe, and \cafe\ are shown Figure \ref{fig:Si Mg Ca LTE-NLTE} with the LTE results from Paper I superimposed (except for Ca which was not measured in Paper I). In addition, the running mean of each trend was computed and is plotted in Figure \ref{fig:Si Mg Ca LTE-NLTE}.

A few differences between LTE and non-LTE can be seen. For \sife, the non-LTE trend is indeed slightly stronger and by looking at its running mean, we see that this trend is only slightly lower than the LTE one. 
For \mgfe, the difference is clearly noticeable; the non-LTE trend is much stronger, showing less scatter, especially for higher metallicities. Moreover, there is a significant offset of $\sim$0.1-0.15 dex between the two trends. 
Finally, for \cafe\ there is almost no difference between the two cases, and only a very slight offset is visible. This latter fact is further discussed in Section \ref{sec:discussion_ca}.\\
By comparing the line-by-line discrepancy values for LTE and non-LTE cases, we see that they are lower in the latter case, which justifies our choice to take into account non-LTE effects, as this appears to improve the accuracy of our spectral analysis. Moreover, we note that using non-LTE decreases the scatter in the [$\alpha$/Fe] versus \feh\ trends.
From now on, we therefore focus on the non-LTE abundances.

\begin{table*}
\caption{$Si$ abundances of the observed stars in the northern fields. The names of stars belonging to our Golden Sample (see Section \ref{sec:golden sample}) are marked in bold face.}\label{table:Si_abund_N}
\begin{tabular}{|l|c|c|c|c|c|c|c|c|c|c|}
\hline
\multirow{2}{*}{Star} & \multicolumn{8}{c|}{\sife\ } & \multirow{2}{*}{$\sigma_{lines}$} & \multirow{2}{*}{\sife$_{mean}$}\\ \cline{2-9}
& 20890.37 & 20908.62 & 20917.11 & 20926.14 & 21056.37 & 21139.76 & 21143.26 & 21204.42 & & \\
\hline
\multicolumn{11}{|l|}{Northern field at $(l,b) = (0^{\circ},+3^{\circ})$}\\
\hline
\textbf{bp3-01} & 0.27       & \noline    & 0.34       & 0.24       & 0.44       & \noline    & 0.21       & \noline    & 0.08    & 0.30    \\
 \textbf{bp3-02} & -0.00      & \noline    & 0.03       & -0.06      & 0.14       & \noline    & \noline    & \noline    & 0.07    & 0.03    \\
 \textbf{bp3-04} & 0.38       & \noline    & \noline    & 0.58       & \noline    & \noline    & \noline    & \noline    & 0.10    & 0.24    \\
 \textbf{bp3-05} & -0.21      & \noline    & -0.42      & -0.28      & -0.21      & \noline    & \noline    & \noline    & 0.09    & -0.22   \\
 bp3-06          & 0.41       & \noline    & 0.59       & \noline    & \noline    & \noline    & \noline    & \noline    & 0.09    & 0.50    \\
 \textbf{bp3-07} & \noline    & -0.10      & -0.15      & -0.24      & -0.05      & \noline    & -0.25      & -0.30      & 0.09    & -0.18   \\
 \textbf{bp3-08} & \noline    & 0.35       & 0.09       & \noline    & 0.09       & \noline    & \noline    & \noline    & 0.12    & 0.11    \\
 \textbf{bp3-10} & 0.03       & \noline    & -0.06      & -0.11      & 0.07       & \noline    & \noline    & -0.14      & 0.08    & -0.04   \\
 \textbf{bp3-11} & 0.07       & \noline    & 0.43       & 0.28       & \noline    & \noline    & \noline    & \noline    & 0.15    & 0.20    \\
 \textbf{bp3-12} & 0.20       & \noline    & 0.23       & 0.22       & 0.20       & \noline    & \noline    & \noline    & 0.01    & 0.14    \\
 \textbf{bp3-13} & 0.01       & \noline    & \noline    & \noline    & -0.03      & \noline    & \noline    & \noline    & 0.02    & -0.01   \\
 \textbf{bp3-14} & 0.03       & \noline    & \noline    & \noline    & -0.03      & \noline    & \noline    & \noline    & 0.03    & 0.00    \\
 \textbf{bp3-15} & 0.33       & \noline    & 0.13       & \noline    & \noline    & \noline    & 0.21       & \noline    & 0.08    & 0.22    \\
 \textbf{bp3-16} & 0.09       & \noline    & 0.47       & \noline    & 0.08       & 0.16       & \noline    & \noline    & 0.16    & 0.20    \\
 \textbf{bp3-17} & \noline    & \noline    & 0.07       & \noline    & 0.02       & \noline    & \noline    & \noline    & 0.02    & 0.03    \\
\hline
 \multicolumn{11}{|l|}{Northern field at $(l,b) = (0^{\circ},+2^{\circ})$}\\
\hline
 \textbf{bp2-01} & -0.10      & \noline    & -0.21      & -0.07      & 0.06       & \noline    & \noline    & \noline    & 0.10    & -0.08   \\
 bp2-02          & 0.23       & 0.76       & \noline    & 0.57       & 0.88       & \noline    & \noline    & \noline    & 0.25    & 0.61    \\
 \textbf{bp2-03} & 0.01       & \noline    & 0.46       & 0.05       & 0.06       & \noline    & \noline    & \noline    & 0.18    & 0.14    \\
 \textbf{bp2-04} & 0.28       & \noline    & 0.29       & 0.25       & 0.37       & \noline    & \noline    & \noline    & 0.04    & 0.30    \\
 \textbf{bp2-05} & 0.30       & \noline    & 0.44       & 0.39       & 0.40       & \noline    & \noline    & \noline    & 0.05    & 0.26    \\
 bp2-06          & \noline    & \noline    & \noline    & \noline    & \noline    & \noline    & \noline    & \noline    & \noline & \noline \\
 \textbf{bp2-07} & 0.05       & \noline    & 0.24       & 0.02       & 0.26       & 0.12       & \noline    & \noline    & 0.10    & 0.14    \\
 \textbf{bp2-08} & 0.10       & 0.99       & 0.62       & -0.03      & 0.27       & \noline    & \noline    & \noline    & 0.37    & 0.39    \\
 \textbf{bp2-09} & 0.17       & \noline    & 0.45       & 0.09       & 0.35       & \noline    & \noline    & \noline    & 0.14    & 0.21    \\
 \textbf{bp2-10} & 0.31       & \noline    & 0.24       & -0.02      & \noline    & \noline    & \noline    & 0.02       & 0.14    & 0.14    \\
 \textbf{bp2-11} & 0.10       & \noline    & 0.39       & 0.08       & -0.09      & \noline    & \noline    & \noline    & 0.17    & 0.12    \\
 \textbf{bp2-12} & 0.41       & 0.46       & 0.50       & 0.48       & 0.49       & 0.26       & 0.53       & 0.18       & 0.12    & 0.41    \\
 \textbf{bp2-13} & 0.11       & \noline    & 0.33       & 0.43       & 0.50       & 0.16       & 0.21       & \noline    & 0.14    & 0.29    \\
 \textbf{bp2-14} & 0.05       & 0.52       & 0.40       & 0.15       & 0.16       & \noline    & \noline    & \noline    & 0.18    & 0.26    \\
 \textbf{bp2-15} & 0.37       & \noline    & 0.32       & \noline    & 0.51       & 0.35       & 0.31       & \noline    & 0.07    & 0.37    \\
\hline
 \multicolumn{11}{|l|}{Northern field at $(l,b) = (0^{\circ},+1^{\circ})$}\\
\hline
 bp1-01          & -0.15      & 0.58       & 0.32       & \noline    & -0.17      & \noline    & \noline    & \noline    & 0.32    & 0.14    \\
 \textbf{bp1-02} & 0.60       & \noline    & 0.52       & 0.32       & 0.52       & 0.61       & 0.49       & \noline    & 0.10    & 0.51    \\
 \textbf{bp1-03} & \noline    & \noline    & 0.14       & -0.06      & -0.12      & \noline    & -0.08      & \noline    & 0.10    & -0.03   \\
 \textbf{bp1-04} & -0.04      & 0.02       & 0.10       & \noline    & 0.02       & -0.04      & 0.03       & \noline    & 0.05    & 0.02    \\
 \textbf{bp1-05} & -0.16      & 0.12       & 0.04       & -0.27      & -0.14      & \noline    & 0.10       & \noline    & 0.14    & -0.05   \\
 \textbf{bp1-06} & 0.17       & \noline    & 0.20       & 0.65       & 0.29       & \noline    & \noline    & \noline    & 0.19    & 0.33    \\
 \textbf{bp1-07} & 0.31       & \noline    & 0.43       & 0.40       & 0.51       & 0.39       & \noline    & 0.41       & 0.06    & 0.41    \\
 \textbf{bp1-08} & 0.20       & \noline    & 0.20       & 0.15       & 0.33       & \noline    & \noline    & \noline    & 0.06    & 0.15    \\
 \textbf{bp1-09} & -0.13      & \noline    & -0.10      & \noline    & -0.43      & \noline    & \noline    & \noline    & 0.15    & -0.22   \\
 \textbf{bp1-10} & 0.03       & \noline    & 0.22       & 0.04       & 0.41       & \noline    & \noline    & \noline    & 0.16    & 0.12    \\
 \textbf{bp1-11} & 0.29       & \noline    & 0.20       & \noline    & 0.34       & \noline    & 0.10       & \noline    & 0.09    & 0.23    \\
 \textbf{bp1-12} & 0.02       & 0.07       & \noline    & 0.08       & 0.10       & \noline    & \noline    & \noline    & 0.03    & 0.07    \\
 \textbf{bp1-13} & 0.06       & \noline    & 0.24       & \noline    & 0.02       & \noline    & 0.07       & -0.08      & 0.11    & 0.06    \\
 \textbf{bp1-14} & \noline    & \noline    & 0.07       & 0.06       & 0.32       & \noline    & \noline    & \noline    & 0.12    & 0.07    \\
\hline
\end{tabular}
\end{table*}

\begin{table*}
\caption{Si abundances of the observed stars in the southern fields. The names of stars belonging to our Golden Sample (see Section \ref{sec:golden sample}) are marked in bold face.}\label{table:Si_abund_S}
\begin{tabular}{|l|c|c|c|c|c|c|c|c|c|}
\hline
\multirow{2}{*}{Star} & \multicolumn{7}{c|}{\sife\ } & \multirow{2}{*}{$\sigma_{lines}$} & \multirow{2}{*}{\sife$_{mean}$}\\\cline{2-8}
& 20801.41 & 20804.20 & 20890.37 & 20908.62 & 20917.11 & 20926.14 & 21056.37 &  & \\
\hline
\multicolumn{10}{|l|}{Galactic Centre field at $(l,b) = (0^{\circ},0^{\circ})$}\\
\hline
 GC1             & \noline    & \noline    & \noline    & \noline    & 0.17       & \noline    & 0.59       &   0.21 &   0.38 \\
 GC20            & \noline    & \noline    & 0.50       & \noline    & 0.16       & \noline    & -0.26      &   0.31 &   0.14 \\
 \textbf{GC22}   & 0.01       & \noline    & -0.04      & \noline    & 0.07       & \noline    & \noline    &   0.05 &   0.02 \\
 GC25            & \noline    & \noline    & -0.29      & \noline    & -0.34      & -0.37      & -0.22      &   0.06 &  -0.31 \\
 GC27            & 0.27       & -0.04      & -0.24      & \noline    & 0.71       & -0.26      & -0.27      &   0.36 &   0.03 \\
 GC28            & \noline    & -0.32      & -0.23      & \noline    & -0.53      & -0.36      & -0.08      &   0.15 &  -0.3  \\
 \textbf{GC29}   & \noline    & \noline    & \noline    & \noline    & -0.51      & \noline    & 0.18       &   0.34 &  -0.16 \\
 GC37            & \noline    & -0.52      & -0.14      & \noline    & 0.04       & \noline    & 0.09       &   0.24 &  -0.13 \\
 \textbf{GC44}   & \noline    & -0.39      & -0.12      & \noline    & -0.46      & -0.33      & -0.43      &   0.12 &  -0.35 \\
\hline
 \multicolumn{10}{|l|}{Southern field at $(l,b) = (0^{\circ},-1^{\circ})$}\\
\hline
  bm1-06          & \noline    & \noline    & \noline    & \noline    & -0.01      & -0.18      & -0.02      &   0.08 &  -0.07 \\
 \textbf{bm1-07} & \noline    & \noline    & 0.11       & \noline    & 0.14       & \noline    & -0.04      &   0.08 &   0.07 \\
 \textbf{bm1-08} & \noline    & -0.26      & -0.14      & \noline    & -0.23      & \noline    & -0.13      &   0.06 &  -0.19 \\
 \textbf{bm1-10} & \noline    & -0.03      & 0.10       & \noline    & 0.09       & \noline    & 0.06       &   0.05 &   0.05 \\
 \textbf{bm1-11} & \noline    & \noline    & \noline    & \noline    & 0.16       & -0.19      & 0.05       &   0.14 &   0.01 \\
 bm1-13          & \noline    & 0.20       & 0.27       & \noline    & 0.32       & \noline    & 0.52       &   0.12 &   0.32 \\
 bm1-17          & \noline    & 0.45       & 0.23       & \noline    & \noline    & \noline    & \noline    &   0.11 &   0.34 \\
 bm1-18          & \noline    & \noline    & -0.29      & \noline    & 0.00       & -0.11      & 0.08       &   0.14 &  -0.08 \\
 bm1-19          & \noline    & -0.05      & \noline    & 0.26       & -0.09      & -0.01      & 0.47       &   0.21 &   0.12 \\
\hline
 \multicolumn{10}{|l|}{Southern field at $(l,b) = (0^{\circ},-2^{\circ})$}\\
\hline
 \textbf{bm2-01} & \noline    & -0.09      & 0.16       & \noline    & 0.13       & 0.03       & \noline    &   0.1  &   0.06 \\
 \textbf{bm2-02} & \noline    & 0.07       & 0.18       & \noline    & 0.21       & 0.20       & 0.37       &   0.1  &   0.21 \\
 \textbf{bm2-03} & \noline    & 0.03       & 0.05       & \noline    & 0.11       & \noline    & 0.20       &   0.07 &   0.1  \\
 \textbf{bm2-05} & \noline    & \noline    & -0.18      & \noline    & 0.40       & \noline    & -0.04      &   0.25 &   0.06 \\
 \textbf{bm2-06} & \noline    & 0.19       & 0.39       & \noline    & 0.18       & 0.31       & 0.28       &   0.08 &   0.27 \\
 \textbf{bm2-11} & \noline    & 0.33       & 0.20       & \noline    & 0.33       & 0.35       & 0.54       &   0.11 &   0.35 \\
 \textbf{bm2-12} & \noline    & \noline    & -0.06      & \noline    & -0.04      & -0.21      & \noline    &   0.08 &  -0.1  \\
 \textbf{bm2-13} & \noline    & 0.04       & \noline    & \noline    & 0.41       & \noline    & 0.20       &   0.15 &   0.22 \\
 \textbf{bm2-15} & \noline    & \noline    & \noline    & \noline    & -0.05      & \noline    & -0.16      &   0.06 &  -0.1  \\
 \textbf{bm2-16} & \noline    & -0.13      & \noline    & \noline    & 0.22       & \noline    & 0.27       &   0.18 &   0.12 \\
\hline
\end{tabular}
\end{table*}

\begin{table*}
\caption{Mg and Ca abundances of the observed stars in the northern fields. The names of stars belonging to our Golden Sample (see Section \ref{sec:golden sample}) are marked in bold face.}\label{table:Mg_Ca_abund_N}
\begin{tabular}{|l|c||c|c|c|c|c|c|c|}
\hline
\multirow{2}{*}{Star}  & \multicolumn{1}{c|}{\mgfe$_{mean}$} & \multicolumn{5}{c|}{\cafe} & \multirow{2}{*}{$\sigma_{lines}$} & \multirow{2}{*}{\cafe$_{mean}$} \\\cline{2-7}
& 21059.76 + triplet  & 20937.90   & 20956.39   & 20962.57   & 20972.53   & 20973.38 & &\\
\hline
 \multicolumn{9}{|l|}{Northern field at $(l,b) = (0^{\circ},+3^{\circ})$}\\
\hline
\textbf{bp3-01} & 0.19       & 0.38       & 0.36       & 0.30       & \noline    & \noline    & 0.02    & 0.35    \\
 \textbf{bp3-02} & -0.19      & 0.30       & \noline    & 0.36       & \noline    & \noline    & 0.03    & 0.33    \\
 \textbf{bp3-04} & 0.18       & \noline    & \noline    & \noline    & \noline    & \noline    & \noline & \noline \\
 \textbf{bp3-05} & -0.38      & 0.09       & 0.39       & 0.25       & \noline    & \noline    & 0.09    & 0.24    \\
 bp3-06          & 0.27       & 0.51       & \noline    & 0.23       & \noline    & \noline    & 0.14    & 0.37    \\
 \textbf{bp3-07} & -0.29      & 0.40       & 0.10       & \noline    & 0.14       & \noline    & 0.09    & 0.22    \\
 \textbf{bp3-08} & -0.21      & \noline    & \noline    & \noline    & \noline    & \noline    & \noline & \noline \\
 \textbf{bp3-10} & -0.03      & \noline    & 0.34       & 0.26       & \noline    & 0.18       & 0.04    & 0.26    \\
 \textbf{bp3-11} & -0.14      & 0.44       & 0.16       & 0.23       & \noline    & \noline    & 0.09    & 0.28    \\
 \textbf{bp3-12} & -0.10      & 0.38       & \noline    & 0.37       & \noline    & \noline    & 0.01    & 0.37    \\
 \textbf{bp3-13} & -0.15      & 0.55       & 0.07       & \noline    & \noline    & \noline    & 0.24    & 0.31    \\
 \textbf{bp3-14} & -0.10      & 0.36       & 0.09       & \noline    & \noline    & \noline    & 0.13    & 0.22    \\
 \textbf{bp3-15} & 0.16       & 0.23       & \noline    & \noline    & \noline    & \noline    & \noline & 0.23    \\
 \textbf{bp3-16} & 0.08       & 0.54       & \noline    & \noline    & \noline    & \noline    & \noline & 0.54    \\
 \textbf{bp3-17} & -0.09      & 0.30       & 0.33       & \noline    & \noline    & \noline    & 0.02    & 0.32    \\
 \hline
 \multicolumn{9}{|l|}{Northern field at $(l,b) = (0^{\circ},+2^{\circ})$}\\
\hline
 \textbf{bp2-01} & -0.32      & \noline    & \noline    & \noline    & \noline    & \noline    & \noline & \noline \\
 bp2-02          & -0.09      & 0.11       & \noline    & 0.18       & \noline    & \noline    & 0.03    & 0.15    \\
 \textbf{bp2-03} & -0.22      & 0.17       & \noline    & \noline    & \noline    & \noline    & \noline & 0.17    \\
 \textbf{bp2-04} & 0.05       & 0.55       & 0.52       & 0.50       & \noline    & \noline    & 0.01    & 0.52    \\
 \textbf{bp2-05} & 0.18       & 0.03       & \noline    & \noline    & 0.16       & \noline    & 0.07    & 0.10    \\
 bp2-06          & \noline    & \noline    & \noline    & \noline    & \noline    & \noline    & \noline & \noline \\
 \textbf{bp2-07} & 0.09       & \noline    & \noline    & \noline    & \noline    & \noline    & \noline & \noline \\
 \textbf{bp2-08} & -0.12      & 0.56       & 0.65       & 0.46       & \noline    & \noline    & 0.05    & 0.55    \\
 \textbf{bp2-09} & 0.11       & \noline    & \noline    & 0.29       & \noline    & \noline    & \noline & 0.29    \\
 \textbf{bp2-10} & -0.28      & 0.23       & \noline    & 0.06       & \noline    & \noline    & 0.09    & 0.15    \\
 \textbf{bp2-11} & -0.17      & 0.38       & 0.15       & 0.05       & \noline    & \noline    & 0.10    & 0.19    \\
 \textbf{bp2-12} & 0.25       & 0.57       & \noline    & \noline    & \noline    & \noline    & \noline & 0.57    \\
 \textbf{bp2-13} & 0.15       & \noline    & \noline    & \noline    & \noline    & \noline    & \noline & \noline \\
 \textbf{bp2-14} & -0.00      & \noline    & \noline    & \noline    & \noline    & \noline    & \noline & \noline \\
 \textbf{bp2-15} & 0.34       & 0.22       & \noline    & \noline    & \noline    & \noline    & \noline & 0.22    \\
 \hline
 \multicolumn{9}{|l|}{Northern field at $(l,b) = (0^{\circ},+1^{\circ})$}\\
\hline
 bp1-01          & -0.32      & 0.30       & 0.58       & 0.31       & \noline    & \noline    & 0.09    & 0.40    \\
 \textbf{bp1-02} & 0.46       & 0.64       & \noline    & 0.77       & \noline    & \noline    & 0.07    & 0.70    \\
 \textbf{bp1-03} & -0.41      & \noline    & 0.38       & 0.19       & 0.32       & \noline    & 0.06    & 0.30    \\
 \textbf{bp1-04} & -0.15      & 0.20       & 0.16       & 0.17       & \noline    & \noline    & 0.01    & 0.18    \\
 \textbf{bp1-05} & -0.25      & 0.09       & 0.27       & 0.22       & \noline    & \noline    & 0.05    & 0.19    \\
 \textbf{bp1-06} & -0.09      & 0.50       & 0.38       & \noline    & \noline    & \noline    & 0.06    & 0.44    \\
 \textbf{bp1-07} & 0.32       & 0.03       & \noline    & 0.55       & \noline    & \noline    & 0.26    & 0.29    \\
 \textbf{bp1-08} & 0.12       & \noline    & \noline    & 0.04       & 0.13       & \noline    & 0.04    & 0.09    \\
 \textbf{bp1-09} & -0.42      & -0.25      & 0.01       & -0.19      & -0.18      & -0.11      & 0.04    & -0.15   \\
 \textbf{bp1-10} & -0.15      & 0.20       & \noline    & \noline    & \noline    & \noline    & \noline & 0.20    \\
 \textbf{bp1-11} & 0.01       & \noline    & \noline    & \noline    & \noline    & \noline    & \noline & \noline \\
 \textbf{bp1-12} & -0.09      & 0.29       & 0.29       & 0.20       & \noline    & \noline    & 0.03    & 0.26    \\
 \textbf{bp1-13} & 0.01       & \noline    & \noline    & \noline    & \noline    & \noline    & \noline & \noline \\
 \textbf{bp1-14} & -0.02      & -0.28      & \noline    & 0.17       & \noline    & \noline    & 0.22    & -0.06   \\
\hline
\end{tabular}
\end{table*}

\begin{table*}
\caption{$Mg$ and $Ca$ abundances of the observed stars in the southern fields. The names of stars belonging to our Golden Sample (see Section \ref{sec:golden sample}) are marked in bold face.}\label{table:Mg_Ca_abund_S}
\begin{tabular}{|l|c||c|c|c|c|c|c|c|}
\hline
\multirow{2}{*}{Star}  & \multicolumn{1}{c|}{\mgfe$_{mean}$} & \multicolumn{5}{c|}{\cafe} & \multirow{2}{*}{$\sigma_{lines}$} & \multirow{2}{*}{\cafe$_{mean}$} \\\cline{2-7}
 & 21059.76 + triplet   &  20937.90   & 20956.39   & 20962.57   & 20972.53   & 20973.38   &  &    \\
\hline
\multicolumn{9}{|l|}{Galactic Centre field at $(l,b) = (0^{\circ},0^{\circ})$}\\
\hline
  GC1             &      -0.1  & 0.20       & 0.26       & \noline    & \noline    & \noline    & 0.03    & 0.23    \\
 GC20            &      -0.25 & 0.74       & 0.46       & 0.40       & 0.28       & 0.22       & 0.09    & 0.42    \\
 \textbf{GC22}   &      -0.28 & 0.40       & 0.37       & 0.20       & \noline    & \noline    & 0.06    & 0.32    \\
 GC25            &      -0.36 & \noline    & 0.57       & 0.38       & \noline    & \noline    & 0.10    & 0.48    \\
 GC27            &      -0.21 & 0.13       & \noline    & 0.27       & \noline    & \noline    & 0.07    & 0.20    \\
 GC28            &      -0.25 & 0.36       & 0.37       & 0.19       & 0.13       & 0.23       & 0.05    & 0.26    \\
 \textbf{GC29}   &      -0.16 & \noline    & \noline    & \noline    & \noline    & \noline    & \noline & \noline \\
 GC37            &      -0.45 & -0.19      & 0.08       & \noline    & \noline    & \noline    & 0.14    & -0.06   \\
 \textbf{GC44}   &      -0.36 & \noline    & \noline    & \noline    & \noline    & \noline    & \noline & \noline \\
\hline
\multicolumn{9}{|l|}{Southern field at $(l,b) = (0^{\circ},-1^{\circ})$}\\
\hline
 bm1-06          &      -0.12 & 0.68       & 0.32       & 0.31       & \noline    & \noline    & 0.12    & 0.43    \\
 \textbf{bm1-07} &      -0.11 & 0.19       & 0.17       & \noline    & \noline    & \noline    & 0.01    & 0.18    \\
 \textbf{bm1-08} &      -0.25 & \noline    & 0.34       & 0.29       & -0.12      & 0.01       & 0.11    & 0.13    \\
 \textbf{bm1-10} &       0.02 & \noline    & \noline    & \noline    & \noline    & \noline    & \noline & \noline \\
 \textbf{bm1-11} &      -0.27 & 0.24       & 0.58       & \noline    & \noline    & 0.24       & 0.11    & 0.35    \\
 bm1-13          &       0.4  & \noline    & \noline    & \noline    & \noline    & \noline    & \noline & \noline \\
 bm1-17          &       0.54 & \noline    & \noline    & \noline    & \noline    & \noline    & \noline & \noline \\
 bm1-18          &      -0.14 & -0.61      & \noline    & \noline    & 0.10       & \noline    & 0.35    & -0.26   \\
 bm1-19          &      -0.14 & -0.64      & 0.32       & \noline    & \noline    & 0.25       & 0.31    & -0.02   \\
\hline
\multicolumn{9}{|l|}{Southern field at $(l,b) = (0^{\circ},-2^{\circ})$}\\
\hline
\textbf{bm2-01} &       0.04 & -0.20      & 0.17       & \noline    & \noline    & \noline    & 0.19    & -0.01   \\
 \textbf{bm2-02} &       0.13 & 0.36       & 0.26       & 0.28       & 0.21       & 0.19       & 0.03    & 0.26    \\
 \textbf{bm2-03} &      -0.04 & 0.19       & \noline    & 0.18       & -0.03      & 0.10       & 0.05    & 0.11    \\
 \textbf{bm2-05} &      -0.08 & 0.31       & 0.36       & \noline    & \noline    & \noline    & 0.03    & 0.33    \\
 \textbf{bm2-06} &       0.29 & \noline    & \noline    & 0.43       & \noline    & \noline    & \noline & 0.43    \\
 \textbf{bm2-11} &       0.2  & 0.21       & \noline    & 0.19       & \noline    & \noline    & 0.01    & 0.20    \\
 \textbf{bm2-12} &      -0.31 & -0.05      & 0.31       & 0.24       & 0.15       & \noline    & 0.08    & 0.16    \\
 \textbf{bm2-13} &       0.11 & 0.16       & -0.16      & \noline    & \noline    & \noline    & 0.16    & 0.00    \\
 \textbf{bm2-15} &      -0.27 & 0.16       & 0.18       & 0.12       & \noline    & 0.07       & 0.02    & 0.13    \\
 \textbf{bm2-16} &      -0.01 & 0.14       & \noline    & 0.14       & 0.09       & 0.22       & 0.03    & 0.15    \\
\hline
\end{tabular}
\end{table*}

\subsubsection{Silicon and magnesium}
\label{sec:results:simg}
 
\begin{figure*}[ht!]
   \centering
   \includegraphics[width=\hsize]{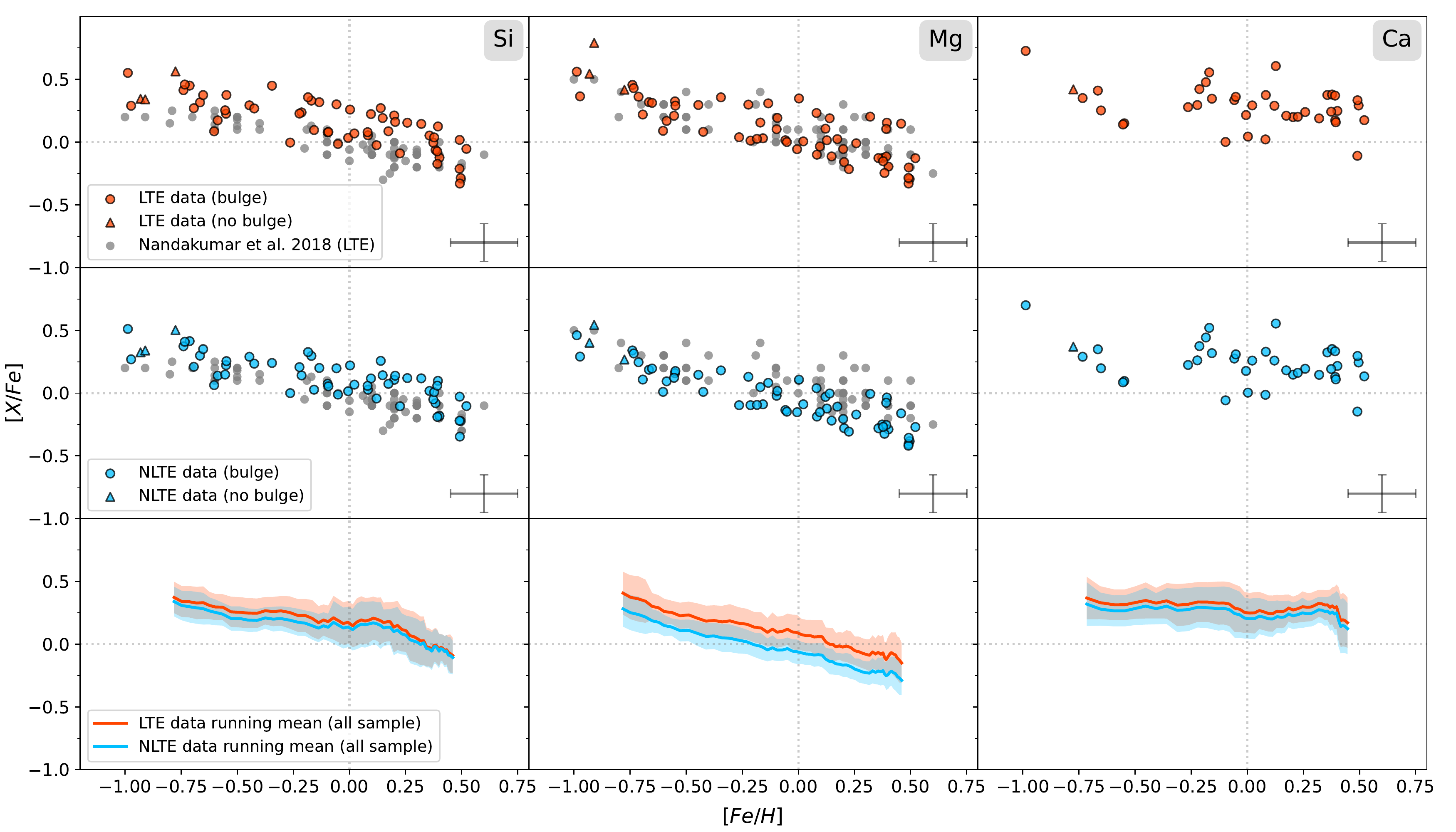}
      \caption{Comparison of our \sife, \mgfe,\ and \cafe\ versus \feh\ trends in LTE and non-LTE. The comparison sample for Si and Mg is from \citet{nandakumar:18} and is shown in grey. The upper row shows our trends in red while assuming LTE in the line formation. The middle row shows our trends in blue when assuming non-LTE instead. The lower row shows the running mean of our trends with and without LTE in order to demonstrate the differences more clearly.}
         \label{fig:Si Mg Ca LTE-NLTE}
\end{figure*}

In Fig.\ref{fig:Si Mg Ca LTE-NLTE} we show our new \sife\ and \mgfe\ trends as function of metallicity compared to the ones derived in Paper I. The uncertainties are the same for the old and the new abundances: 0.15 dex. Besides the fact that our \sife\ and \mgfe\ trends follow a typical $\alpha$-element trend, we observe that there is no plateau at supersolar metallicities for either  silicon or magnesium; our trends keep going down. Also, our Si abundances are slightly higher (+0.14 dex in average) and our Mg abundances slightly lower (-0.14 dex in average) than the results from Paper I. Compared to the latter, where \sife\ results were showing a lower dispersion than \mgfe,\, we get the opposite in our study. This is explained by the fact that the Mg lines used are more sensitive to non-LTE effects than the Si lines (see Figure \ref{fig:Si Mg Ca LTE-NLTE}).

\subsubsection{Calcium}

The determination of calcium abundances for our inner bulge stars is new compared to Paper I and is therefore one of the major novelties of this study. As mentioned in Section \ref{sec:spectral analysis}, only the few stars with satisfactory Ca lines have a calcium abundance estimation (see final values in Table \ref{fig:spectra all si mg ca lines}). We can see that our Ca trend is more dispersed (which could be simply due to the lack of stars) and higher values are seen for metal-rich stars, which is different from our Si and Mg trends.

\subsection{Distances}
\label{sec:distances}
We determined the spectro-photometric distances using the stellar parameters \teff, \logg, and \feh\ \citep{distance1,Schultheis2017} together with the near-IR photometry J, H, and K and associated errors, in order to simultaneously compute the most likely line-of-sight distance and reddening by isochrone fitting using PARSEC isochrones. Here we use the same method as that described in \citet{Rojas2017}, where we consider a set of isochrones spanning ages from 1 to 13 Gyr in steps of 1\,Gyr and metallicities from -2.2 to +0.5\,dex in steps of 0.1 dex. As in  \citet{Rojas2017}, besides computing the distance from the observed stars to the whole set of model stars,  we  calculate weights for each star, which depend on: the evolutionary speed of the model stars; the fact that the number of stars per mass is not uniform, which is given by the IMF; and an exponential weight associated with the distance of the observed stars with respect to the model. The typical uncertainties are on the order of 20\%-40\% but we find no systematic differences in comparison to distances obtained by a Bayesian approach for example (e.g. \citealt{Santiago2016}).  The obtained distances permit us to analyse the bulge membership of our sample; see Section \ref{sec:membership}. We note that we do not calculate the distances of the Galactic centre field because we lack the J magnitudes. From the heliocentric distances, we computed the Galactocentric radial distance $\rm  R_{GC} = \sqrt{(X_{GC}^2 + Y_{GC}^2)} $ where $\rm X_{GC}$
 and $\rm Y_{GC}$ are the cartesian coordinates.
 
\begin{table}
\caption{Distances of the observed stars in the northern fields. The names of stars belonging to our Golden Sample (see Section \ref{sec:golden sample}) are marked in bold face.}\label{table:distances_N}
\begin{tabular}{|l|c|c|c|c|}
\hline
\multicolumn{1}{|l|}{Star} & $R_{GC}$ & Distance error & Population\\
         & [kpc] & [kpc] & \\  
\hline
\multicolumn{4}{|l|}{Northern field at $(l,b) = (0^{\circ},+3^{\circ})$}\\
\hline
 \textbf{bp3-01} &   0.3 &              0.7 & bulge        \\
 \textbf{bp3-02} &   0.4 &              1.0 & bulge        \\
 \textbf{bp3-04} &   1.7 &              1.1 & bulge        \\
 \textbf{bp3-05} &   2.2 &              0.8 & bulge        \\
 bp3-06          &  11.9 &              0.9 & No bulge     \\
 \textbf{bp3-07} &   1.9 &              0.9 & bulge        \\
 \textbf{bp3-08} &   1.3 &              0.8 & bulge        \\
 \textbf{bp3-10} &   1.1 &              1.0 & bulge        \\
 \textbf{bp3-11} &   0.9 &              0.6 & bulge        \\
 \textbf{bp3-12} &   1.6 &              1.1 & bulge        \\
 \textbf{bp3-13} &   0.4 &              0.9 & bulge        \\
 \textbf{bp3-14} &   1.6 &              0.6 & bulge        \\
 \textbf{bp3-15} &   1.8 &              0.6 & bulge        \\
 \textbf{bp3-16} &   0.2 &              0.9 & bulge        \\
 \textbf{bp3-17} &   1.9 &              0.7 & bulge        \\
\hline
 \multicolumn{4}{|l|}{Northern field at $(l,b) = (0^{\circ},+2^{\circ})$}\\
\hline
 \textbf{bp2-01} &   1.8 &              0.9 & bulge        \\
 bp2-02          &   0.6 &              0.8 & bulge        \\
 \textbf{bp2-03} &   2.0 &              0.7 & bulge        \\
 \textbf{bp2-04} &   2.2 &              0.6 & bulge        \\
 \textbf{bp2-05} &   0.9 &              0.9 & bulge        \\
 bp2-06          &   8.7 &              2.1 & No bulge     \\
 \textbf{bp2-07} &   1.8 &              0.8 & bulge        \\
 \textbf{bp2-08} &   1.4 &              0.7 & bulge        \\
 \textbf{bp2-09} &   1.9 &              1.1 & bulge        \\
 \textbf{bp2-10} &   1.8 &              0.7 & bulge        \\
 \textbf{bp2-11} &   1.1 &              0.7 & bulge        \\
 \textbf{bp2-12} &   0.2 &              1.1 & bulge        \\
 \textbf{bp2-13} &   1.5 &              1.1 & bulge        \\
 \textbf{bp2-14} &   1.6 &              1.3 & bulge        \\
 \textbf{bp2-15} &   1.4 &              0.7 & bulge        \\
\hline
 \multicolumn{4}{|l|}{Northern field at $(l,b) = (0^{\circ},+1^{\circ})$}\\
\hline
 bp1-01          &   1.7 &              0.9 & bulge        \\
 \textbf{bp1-02} &   0.6 &              0.8 & bulge        \\
 \textbf{bp1-03} &   2.1 &              0.9 & bulge        \\
 \textbf{bp1-04} &   2.1 &              0.8 & bulge        \\
 \textbf{bp1-05} &   0.3 &              1.0 & bulge        \\
 \textbf{bp1-06} &   1.0 &              1.0 & bulge        \\
 \textbf{bp1-07} &   0.1 &              0.8 & bulge        \\
 \textbf{bp1-08} &   1.1 &              0.7 & bulge        \\
 \textbf{bp1-09} &   0.1 &              0.9 & bulge        \\
 \textbf{bp1-10} &   1.1 &              0.7 & bulge        \\
 \textbf{bp1-11} &   1.5 &              0.9 & bulge        \\
 \textbf{bp1-12} &   1.8 &              0.9 & bulge        \\
 \textbf{bp1-13} &   1.3 &              0.6 & bulge        \\
 \textbf{bp1-14} &   0.5 &              0.7 & bulge        \\
\hline
\end{tabular}
\end{table}

\begin{table}
\caption{Distances of the observed stars in the southern fields. The names of stars belonging to our Golden Sample (see Section \ref{sec:golden sample}) are marked in bold face.}\label{table:distances_S}
\begin{tabular}{|l|c|c|c|c|}
\hline
\multicolumn{1}{|l|}{Star} & $R_{GC}$ & Distance error & Population\\
         & [kpc] & [kpc] & \\  
\hline
\multicolumn{4}{|l|}{Galactic Centre field at $(l,b) = (0^{\circ},0^{\circ})$}\\
\hline
 GC1             & \noline & \noline          & \noline      \\
 GC20            & \noline & \noline          & \noline      \\
 \textbf{GC22}   & \noline & \noline          & \noline      \\
 GC25            & \noline & \noline          & \noline      \\
 GC27            & \noline & \noline          & \noline      \\
 GC28            & \noline & \noline          & \noline      \\
 \textbf{GC29}   & \noline & \noline          & \noline      \\
 GC37            & \noline & \noline          & \noline      \\
 \textbf{GC44}   & \noline & \noline          & \noline      \\
\hline
 \multicolumn{4}{|l|}{Southern field at $(l,b) = (0^{\circ},-1^{\circ})$}\\
\hline
 bm1-06          &   1.5 &              0.9 & bulge        \\
 \textbf{bm1-07} &   0.2 &              1.1 & bulge        \\
 \textbf{bm1-08} &   1.4 &              0.9 & bulge        \\
 \textbf{bm1-10} &   1.1 &              1.1 & bulge        \\
 \textbf{bm1-11} &   1.8 &              0.8 & bulge        \\
 bm1-13          &   3.9 &              0.4 & No bulge     \\
 bm1-17          &   6.6 &              0.6 & No bulge     \\
 bm1-18          &   0.6 &              1.1 & bulge        \\
 bm1-19          &   1.6 &              1.0 & bulge        \\
\hline
 \multicolumn{4}{|l|}{Southern field at $(l,b) = (0^{\circ},-2^{\circ})$}\\
\hline
 \textbf{bm2-01} &   1.4 &              1.1 & bulge        \\
 \textbf{bm2-02} &   1.0 &              0.8 & bulge        \\
 \textbf{bm2-03} &   1.4 &              1.1 & bulge        \\
 \textbf{bm2-05} &   0.2 &              0.8 & bulge        \\
 \textbf{bm2-06} &   1.8 &              0.8 & bulge        \\
 \textbf{bm2-11} &   0.2 &              1.0 & bulge        \\
 \textbf{bm2-12} &   1.8 &              0.9 & bulge        \\
 \textbf{bm2-13} &   1.1 &              1.0 & bulge        \\
 \textbf{bm2-15} &   1.8 &              0.8 & bulge        \\
 \textbf{bm2-16} &   1.1 &              0.9 & bulge        \\
\hline
\end{tabular}
\end{table}

\subsection{Bulge membership}
\label{sec:membership}
\begin{figure}[ht!]
   \centering
   \includegraphics[width=\hsize]{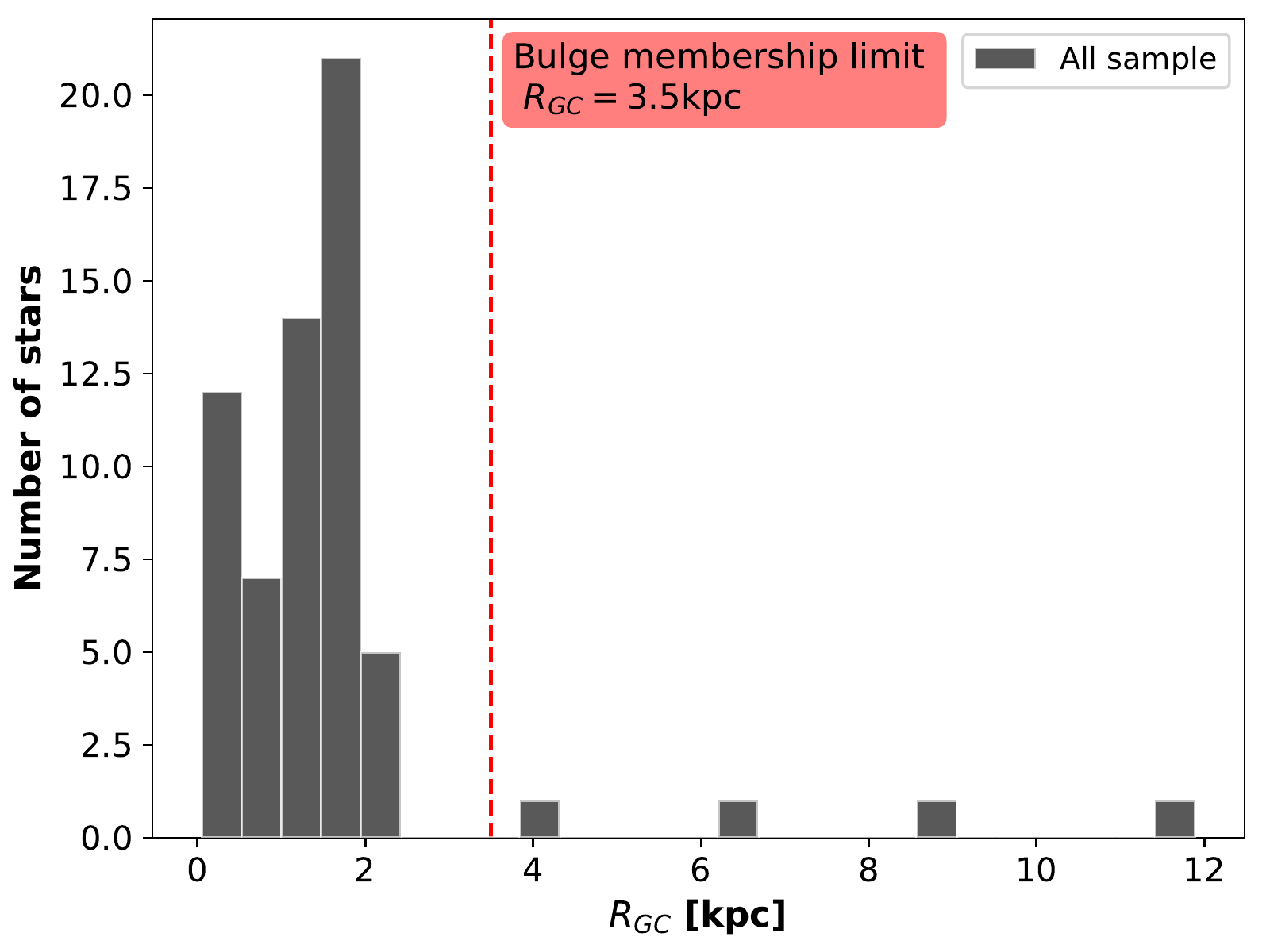}
      \caption{Histogram of the Galactocentric radial distances (computed in Section \ref{sec:distances}) of the stars in our sample for which 
values are available. The typical intrinsic depth of the bulge is indicated by the red dashed line.}
         \label{fig:rgc}
\end{figure}

Figure ~\ref{fig:rgc} displays the distribution of the Galactocentric distances ($\rm R_{GC}$) where we see a peak in the distribution at $\rm R_{GC} \sim 1.5\,kpc$. We also indicate the typical intrinsic depth of the bulge with $\rm R_{GC} \leq 3.5\,kpc$ (see e.g. \citealt{Rojas2017}, \citealt{Rojas2020}, \citealt{Zasowski2019}).
Four stars (bp3-06, bp2-06, bm1-13, bm1-17) show larger distances going up to $\rm R_{GC} \sim 12.0\,kpc,$ indicating that those do not belong to the inner bulge. They all have low metallicities ($\rm [Fe/H] \sim -1.0$) and are enhanced in the $\alpha-$elements and could therefore belong either to the thick disc or  to the halo population. Thus, compared to Paper I where no star was found outside of the bulge, in this study we exclude those four stars from further analysis, resulting in a total sample of 59 stars with determined
distances.

\begin{figure}[ht!]
   \centering
   \includegraphics[width=\hsize]{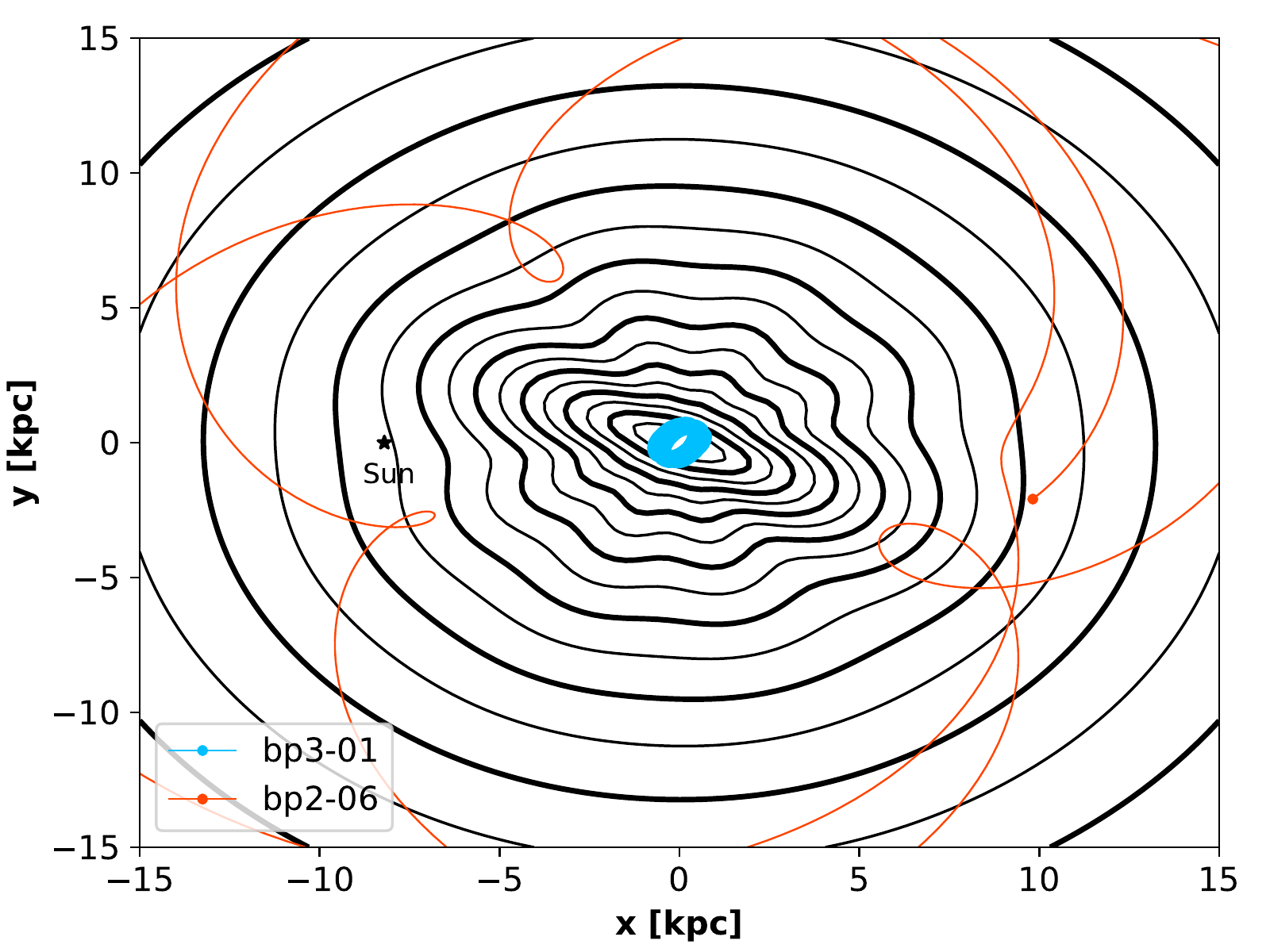}
      \caption{Comparison of the orbits in the Cartesian Galactocentric (x,y) plane of bp3-01 found in the bulge and bp2-06 found outside of the bulge. The black contours show the surface density of the stellar component of our combination of gravitational potentials.}
         \label{fig:orbits}
\end{figure}

 All the stars of our sample were spectrally analysed using the method explained in Section \ref{sec:spectral analysis}, except the star bp2-06. Indeed, due to the very low metallicity of this latter object (the estimation for  bp2-06 from Paper I gives $-1.8$ dex), none of our Fe lines are usable for determining it directly. However, thanks to our dynamical analysis introduced in Section \ref{sec:dynamics} and to our broad Mg lines, we can estimate its metallicity. As explained in Section \ref{sec:distances} and shown in Table \ref{table:distances_N}, the Galactocentric radial distance of bp2-06 is about $8.7$ kpc (using its metallicity estimation from Paper I), which demonstrates that it does not belong to the inner bulge. In addition, its simulated orbit shown in Figure \ref{fig:orbits} supports the latter finding and provides further information about its movements in the Galaxy.

These results lead us to conclude that bp2-06 may be a halo star and therefore it ought to have a typical Mg abundance for a halo star of about $\sim0.4$ dex \citep[see e.g.][]{edvarsson:93}. A magnesium-based metallicity can therefore be estimated by selecting a range of iron abundances \feh\ and corresponding \logg\ and \vmic\ and by fitting our Mg spectral feature for each of these sets of values. We finally obtain a metallicity of about $-2.03$ dex, which allows us to obtain a Mg abundance of about $0.4$ dex for this star. Such a low metallicity, together with its orbits, leads us to reject the notion that this star is in the inner bulge. It is therefore excluded from our inner bulge sample. 

We also checked the derived orbital parameters  for our bulge stars sample, such as apocentric radius ($\rm R_{apo}$) and maximum vertical height ($\rm z_{max}$). Except for the  four  stars with larger distances mentioned above (bp3-06, bp2-06, bm1-13, bm1-17), our sample shows typical orbital parameters for the bulge, with the majority of stars lying within $\rm R_{apo} < 3\,kpc$ and $\rm z_{max} < 1.5\,kpc$ and with high  eccentricities. We do not find any trends in the orbital parameters with metallicity and/or abundances, but a larger sample is clearly needed in order to obtain statistically
significant  results.

\subsection{The Golden Sample}
\label{sec:golden sample}
The objective of this work is to derive the most precise abundance trends as possible for the inner Galactic bulge. In this context, we construct a clean sample, referred to as the Golden Sample, which contains stars belonging to the inner bulge as determined in Section \ref{sec:distances} and that have sufficiently high-quality spectra. To select these stars (i) we require a minimum S/N of 40 (see Section \ref{sec:spectral analysis}) and (ii) as explained in Section \ref{sec:membership}, we exclude stars with $R_{GC}>3.5$ kpc.

Therefore, because of an insufficient S/N, we discarded  the stars bp2-02, bp1-01, GC1, GC20, GC25, GC27, GC28, GC37, bm1-06, bm1-18, and bm1-19 from the abundance analysis. It should be noted that the latter are mainly from the Galactic centre field, which is expected as it corresponds to the innermost region studied here. Similarly, the stars bp3-06, bp2-06, bm1-13, and bm1-17 are excluded because of their excessive distances. From the original sample of 72 stars, our Golden Sample is therefore reduced to 57 stars. The latter are highlighted in bold face in Tables \ref{table:stellarparam_N}, \ref{table:stellarparam_S}, \ref{table:Si_abund_N}, \ref{table:Si_abund_S}, \ref{table:Mg_Ca_abund_N}, \ref{table:Mg_Ca_abund_S}, \ref{table:distances_N}, and \ref{table:distances_S}.

Moreover, we computed the line-by-line discrepancy, $\sigma_{lines}$ (given in Table.\ref{table:Si_abund_N},\ref{table:Si_abund_S},\ref{table:Mg_Ca_abund_N} and \ref{table:Mg_Ca_abund_S}), for every star abundance, which is the standard deviation of the used line abundance values.
Therefore, in addition to the selection made for the construction of our Golden Sample, we require $\sigma_{lines} < 0.3$ for each chemical element. As explained in Section \ref{sec:spectral analysis}, due to the fact that we used only one spectral feature for magnesium, the line-by-line discrepancy is zero for this element. This selection criterion is therefore useful to our analysis, but as its value strongly depends on the number of used lines, caution must be taken when using it; for example, abundances determined with only one line cannot be evaluated by this criterion.


\section{Discussion}
Here, we discuss our results presented in Section \ref{sec:results} and compare them to a local sample of solar neighbourhood stars, for which we used the same lines and spectral analysis method (we refer the reader  to Section \ref{sec:obs_SN} and \ref{sec:analysis_SN} for more details). Moreover, our abundance trends are also compared to the APOGEE inner bulge sample from \citet{Zasowski2019} (hereafter referred to as the APOGEE sample).  The APOGEE sample has been updated with the latest APOGEE DR17 release (\citealt{APOGEEDR17}) covering about 4000 stars with $\rm R_{GC} < 4\,kpc$. Finally, we discuss our abundances in context of a tailored chemical-evolution model for the inner bulge. 

\subsection{Silicon and magnesium}

\begin{figure*}[ht!]
   \centering
   \includegraphics[width=\hsize]{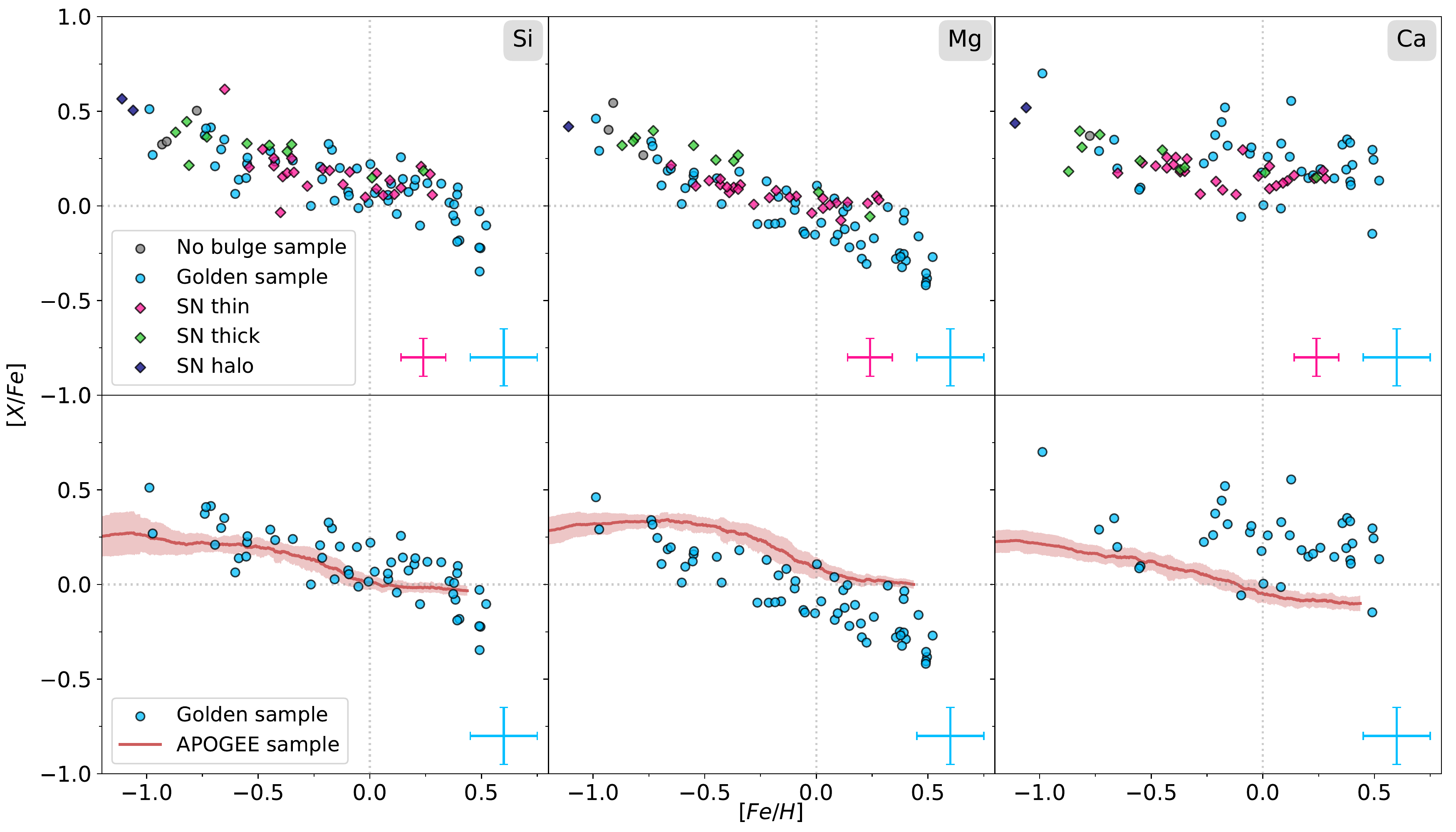}
      \caption{NLTE \sife, \mgfe,\ and \cafe\ vs. \feh\ trends of our Golden Sample of the inner bulge shown in light blue and compared to solar neighbourhood stars in purple (thin disc), green (thick disc), and dark blue (halo) in the {\it upper panel} and compared to the running mean of the updated APOGEE DR17 sample from \cite{Zasowski2019} plotted in red in the {\it lower panel}.}
         \label{fig:sn&zasowski}
\end{figure*}

The NLTE Si and Mg abundance trends for our Golden Sample of the inner bulge are shown in Figure \ref{fig:sn&zasowski}. 
These are compared to our local star sample in the upper panel of the figure. This comparison provides a more direct test to look for differences or similarities
between the populations, as the analyses of the stars are designed to be as similar as possible, thereby minimising any systematic
uncertainties. As indicated in Section \ref{sec:analysis_SN}, even if the local sample has been selected in the most reasonable way possible,  the potential existence of unknown systematic uncertainties in \teff \ should be kept in mind while comparing the two samples. However, a larger scatter is  expected for the inner bulge sample, due to the most obvious and only significant  difference between the analyses, which is the S/Ns. 
In general, we see  that the  Si and Mg trends for the inner bulge and local thin and thick discs are in very good agreement. 

However, our inner bulge stars extend  to higher metallicities. Furthermore, although based on few stars, the local sample shows a tendency to `flatten off' at \feh$>0.25$\,dex. The inner bulge sample, on the other hand, clearly shows a continuous downward trend with increasing metallicity (\cite{johnson2013} report a similar result for several $\alpha$ elements in outer bulge fields; moreover, chemical-evolution models suggest this is indeed to be expected; see Sect.~\ref{sec:CEM}.), albeit with a significant scatter in the abundance ratios. This is true for both the \sife\ and \mgfe\ trends. Our derived Mg abundances show the steepest slope of  decreasing Mg with increasing \feh\ with a mean dispersion of $~0.10$ dex (see Fig.~\ref{fig:Si Mg Ca LTE-NLTE}). More stars and further investigations into the cause of the increased scatter at supersolar metallicities for the inner bulge stars are needed in order to be able to claim any difference in the slopes of the trends for the two populations.

In the lower panel of Fig.~\ref{fig:sn&zasowski}, the Si and Mg trends for our inner bulge stars are compared to those of the inner bulge APOGEE sample from \citet{Zasowski2019}. These data result from different analysis methods and a systematic shift might therefore be expected. Silicon is the $\alpha$-element with the lowest dispersion in APOGEE, which is also seen in other IR studies (\citealt{Rich2005}, \citealt{Schultheis2017}, \citealt{Ryde2010}). We see good agreement
overall, within the uncertainties, except for the highest metallicities. There is, on the other hand, a significant and systematic offset between the data sets for the \mgfe\ trends,  with the metal-rich regime showing the largest differences, namely of up to 0.4\,dex. We note that
Mg  and Si are hailed to be the $\alpha$-elements measured with the greatest precision by APOGEE (see e.g. \citealt{Joensson2018}, \citealt{Schultheis2017}) and Mg does not show signs of any temperature dependency. As this offset is also the case for our local disc sample, and the APOGEE \mgfe\ trend corresponds to what is expected for a bulge/thick disc trend \citep[see e.g.][]{bensby:14}, we believe that our \mgfe\ trend is systematically too low by a few times 0.1 dex.  As noted above, such a shift might be expected in light of the differences in the analysis methods, such as the choice of lines used, whether or not NLTE effects are used, and whether or not those effects are correct in magnitude. Therefore, a differential comparison of the local and the inner bulge stars is very important. 

We also note that a flattening in \sife\ and \mgfe\ trends is seen for the APOGEE sample for metal-rich stars, that is, $\rm [Fe/H] > 0.0$. \citet{Santos-Paral2020} demonstrated that the normalisation procedure has an important impact on the derived abundances and that narrow normalisation windows around the studied lines improves the precision and leads to a decreasing trend in the abundance even at supersolar metallicities. This might be an explanation for the differences.  However, we note that this explanation would not explain this levelling-off tendency for our local sample.

\begin{figure*}[ht!]
   \centering
   \includegraphics[width=\hsize]{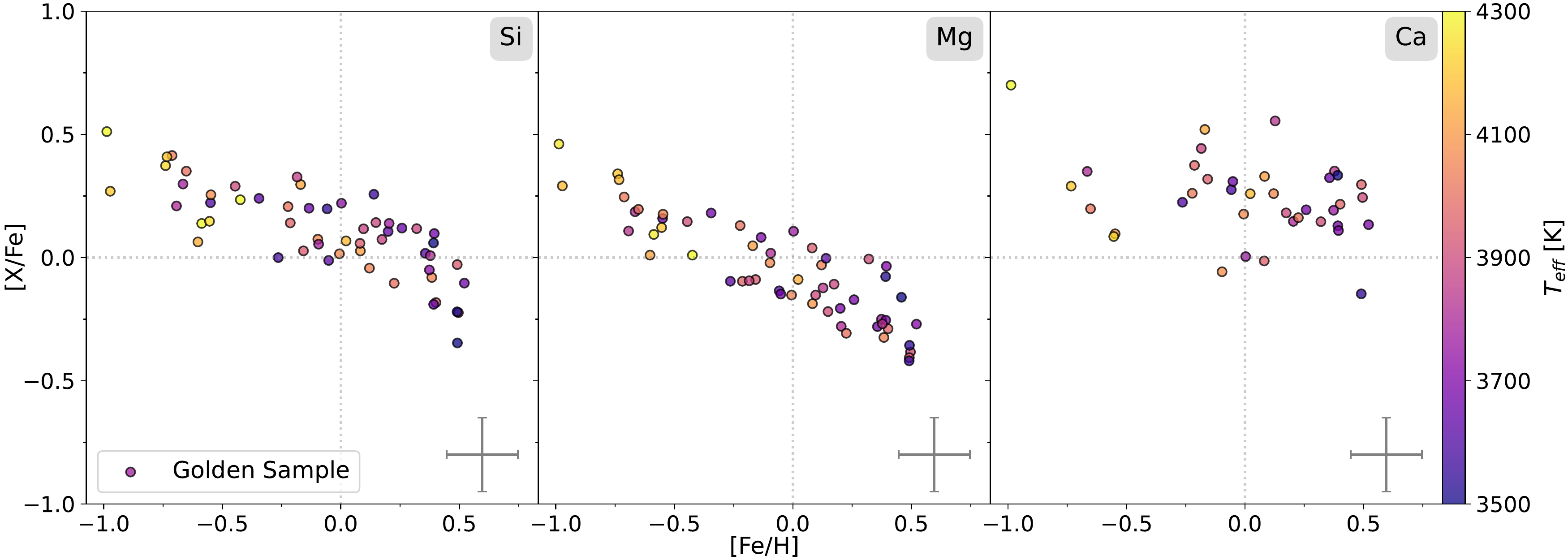}
      \caption{NLTE \sife, \mgfe,\ and \cafe\ vs. \feh\ trends of our Golden Sample of the inner bulge colour coded with \teff\ showing no bias with temperature.}
         \label{fig:trends&teff}
\end{figure*}

The APOGEE sample suffers from some slight temperature dependencies of Si abundance, which is not the case for our Golden Sample, as shown in Figure \ref{fig:trends&teff} (only an expected trend is visible: the warmer stars are more metal poor). This might affect the more metal-rich APOGEE stars, as these tend to be cooler. However, the \sife\ difference in the slope at supersolar metallicities is less obvious when comparing with our inner bulge sample, up to \feh$\sim0.3$. 
As shown in Fig.~\ref{fig:Si Mg Ca LTE-NLTE}, NLTE corrections for Si are also very small, and so this should not cause any differences. 

\subsection{Calcium}
\label{sec:discussion_ca}

The Ca abundances determined for the local solar neighbourhood sample show only slightly larger scatter compared to the local \sife\ and the \mgfe\ trends. However, the \cafe\ trend for our golden sample of the  inner bulge  shows the largest intrinsic dispersion  compared to Mg and Si. This is in part due to problems with the Ca lines, such as the fact that they are situated close to the edge of the detector in the CRIRES setup. This is not the case for the solar neighbourhood sample, which is observed with the IGRINS spectrometer, covering the full K band with no gaps. 
Nevertheless, we see a general agreement in \cafe\ between the solar neighbourhood and the Golden Sample.

A clear offset compared with the APOGEE sample can be seen. \citet{Schultheis2017} noted that, for Baade's window, the Ca abundances of APOGEE are systematically  $\sim$ 0.15\,dex lower than those reported by \cite{Rich2005},\cite{fulbright2007}, and \cite{Gonzalez2011}. The APOGEE trend is also lower compared to what is expected for the bulge/thick disc trend \cite[see e.g.][]{bensby:14}. This shift would put our local sample and APOGEE's trend in closer agreement. 
The NLTE corrections for Ca could also be underestimated for the K band lines, leading to overly high Ca abundances with respect to APOGEE.
However, the Ca NLTE corrections \citep{NLTE} for the H band are  small and the solar disc trends with these NLTE corrections show good agreement with the \cafe\ trends for the disc determined from optical lines \citep{montelius:22}. There is therefore no direct reason to believe that these corrections should be underestimated. As for the other $\alpha$-elements,  here we again see  the flattening in APOGEE for the high-metallicity stars.

\begin{figure*}[ht!]
   \centering
   \includegraphics[width=\hsize]{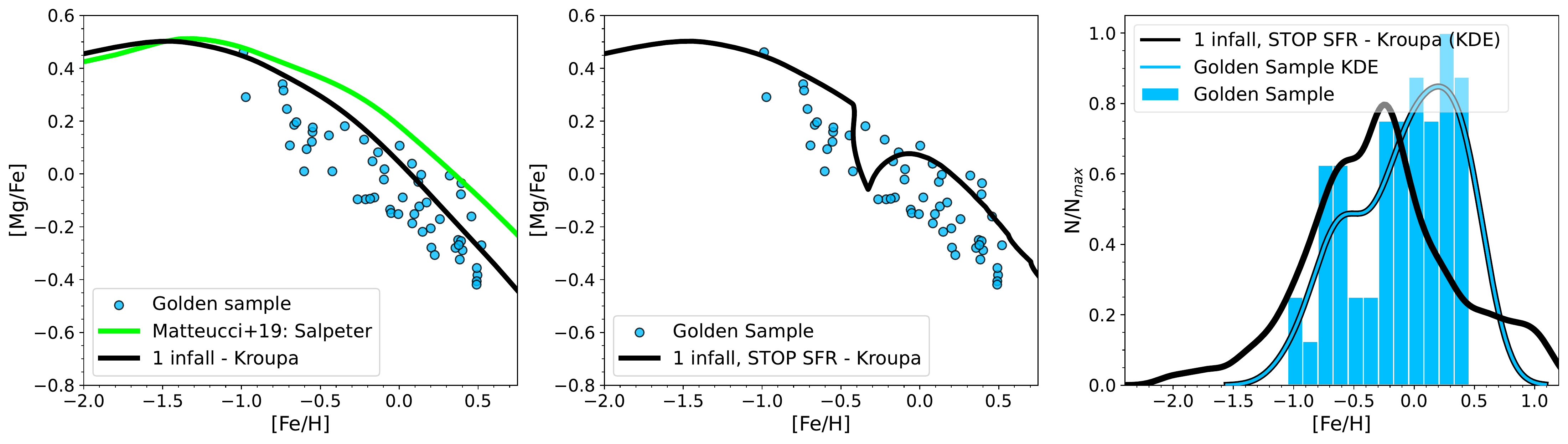}
      \caption{One-infall chemical-evolution model predictions for the Galactic bulge. {\it Left panel:} Effects of different IMF on the one-infall chemical-evolution model. With the green line, we show the model presented  in \citet{matteucci2019} assuming the \citet{salpeter1955} IMF, whereas with the black line we show the same model but with the \citet{kroupa1993} IMF. Blue dots are the \mgfe\ abundances for our Golden Sample.  {\it Middle panel:} [Mg/Fe] vs. [Fe/H] for the one-infall model with the \citet{kroupa1993} IMF, where a 250 Myr pause in star formation has been applied.  {\it Right panel:} Kernel density estimation of the MDF of the same model as in the middle panel (black line) compared with the histogram and KDE of the NLTE stars distribution (blue histogram and line).}
         \label{1infall}
\end{figure*}

\begin{figure*}[ht!]
   \centering
   \includegraphics[width=\hsize]{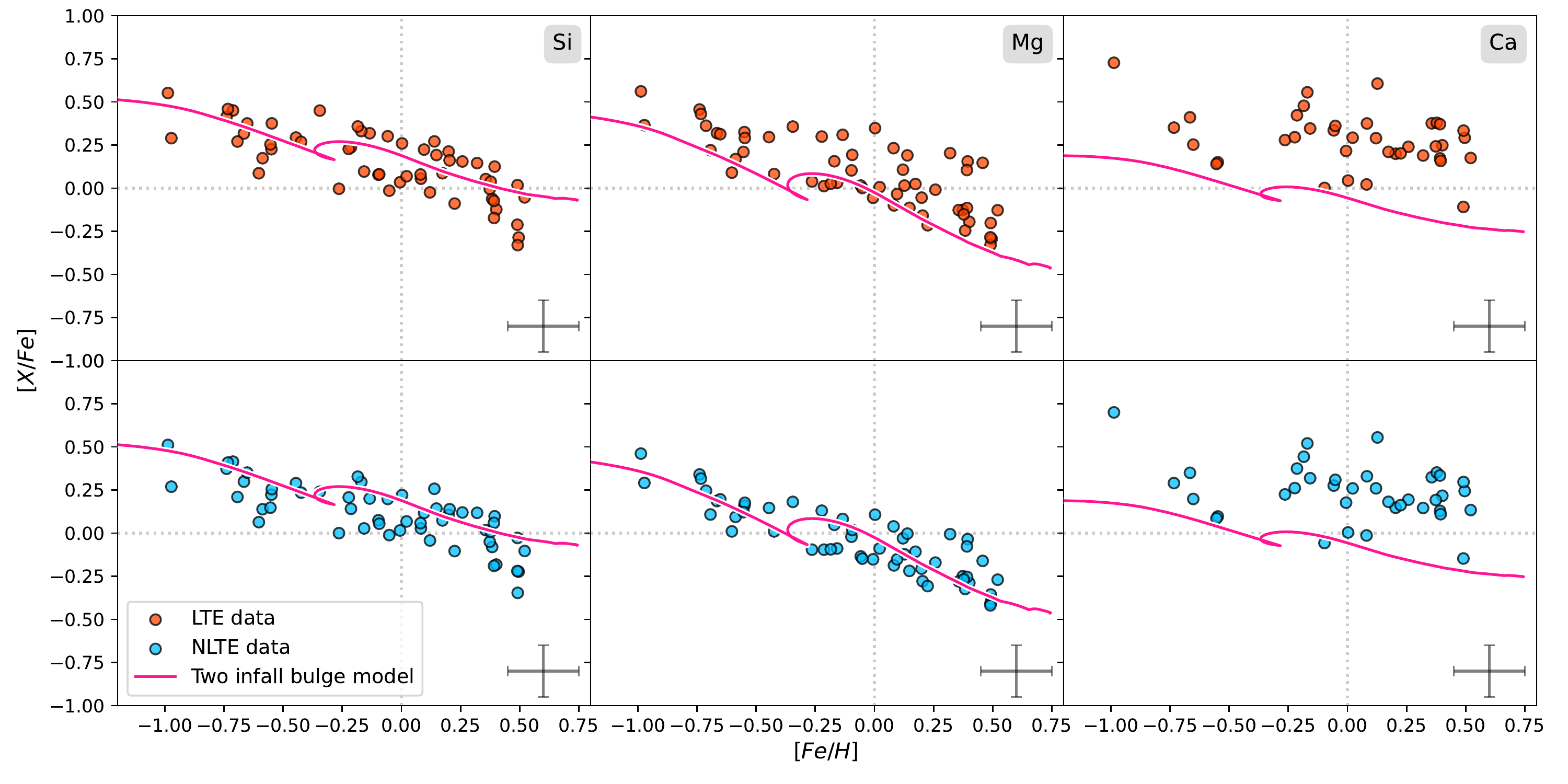}
      \caption{LTE (\textit{upper panel}) and NLTE (\textit{lower panel}) \sife, \mgfe,\ and \cafe\ vs. \feh\ trends of our Golden Sample compared with the two-infall bulge chemical model constructed Section \ref{sec:2-infall}.}
         \label{fig:final trends with model}
\end{figure*}

\subsection{A new  chemical-evolution model of the inner Galactic bulge}
\label{sec:CEM}
In this section,  we present a new chemical-evolution model designed to reproduce the NLTE abundance ratios and metallicity distribution function (MDF) of the inner bulge stars. First, in Section \ref{1infall_sec},  we compare the new  NLTE data for the inner bulge with the chemical-evolution model of \citet{matteucci2019} based on the one-infall scenario designed for the Galactic bulge. Subsequently, in Section \ref{sec:2-infall}, we compare the data to a two-infall model. While such simple approaches to galaxy evolution cannot account for the full complexity of galaxy formation, they have served as useful reference models in the literature for studying the impact of single or multiple episodes of galactic star formation  on the chemical signatures in stellar populations.

\subsubsection{Testing the one-infall chemical-evolution  model }\label{1infall_sec}
In the left panel of Fig. \ref{1infall}, we compare the NLTE [Mg/Fe] versus [Fe/H] data presented in Section \ref{sec:results:simg} with the one-infall chemical-evolution model for the Galactic bulge  of \citet{matteucci2019}.
In this model, the bulge follows a fast gravitational collapse scenario \citep{larson1976} and is supposed to form by accretion of a single fast gas-infall event on a timescale of $\tau$ = 0.1 Gyr. 
The proposed model was capable of reproducing Gaia-ESO bulge stars  of \citet{Rojas2017} adopting  the \citet{salpeter1955} IMF. However, we note from Fig. \ref{1infall}  that the NLTE data presented in this  study are not reproduced by the model. A better agreement is obtained with the  \citet{kroupa1993} IMF. 
As  highlighted in the VINTERGATAN simulation \citep{vintergatan1}, the scatter in  the data with high-[$\alpha$/Fe] abundance ratio values across the entire galaxy is largely set by  the gas -consumption timescales in star forming clouds (see Figure 9 in \citealt{vintergatan3}). Gas with short depletion times shows the highest values of [$\alpha$/Fe] at fixed [Fe/H] values. This is due to supernova-enriched gas being trapped more easily in dense clouds where the depletion time is short. It is worth noticing that in the VINTERGATAN simulation, the scatter in the high-alpha sequence is present in both the in situ formed stars and the accreted stars (see Figure 8 in \citealt{vintergatan3}).

In the middle panel, following \citet{matteucci2019}, in order to reproduce the bimodality in both in the [Mg/Fe] versus [Fe/H] space and in the MDF, we try a 250 Myr  pause  in star formation.
 The chemical-evolution model characterised by a unique  gas-accretion event  coupled with   a pause in star formation is able  to predict the  gap in the [X/Fe] versus [Fe/H]  observed in APOGEE bulge data (middle panel),  but completely fails to properly reproduce the two stellar populations in the MDF (see right panel).

\subsubsection{Inner-bulge stellar populations with the two-infall model}
\label{sec:2-infall}
In the previous section, we show that the classical one-infall chemical-evolution model 
does not seem to be the correct scenario to properly reproduce the chemical properties of the inner bulge.

Here, we   consider the possibility that two gas-accretion events are responsible for the chemical signatures impressed  in the NLTE abundance ratios. The presence of two distinct sequences  in the chemical abundances  in the Galactic  bulge  was pointed out by  \citet{matteucci2019} and  \citet{queiroz2020} based on analyses of APOGEE stars.  In particular, in \citet{queiroz2020},  the  Galactic disc dichotomy in the abundance ratios [$\alpha$/Fe] versus [Fe/H] seemed to extend up to  the innermost regions of the disc up to  the bulge.
The complex nature of the Galactic bulge structures has already been addressed by theoretical models  with the introduction of two different stellar populations.
\citet{samland2003}, by means of a dynamical model, predicted the existence of two bulge populations: one originating  from an early collapse and the other formed late in the bar through  secular evolution \citep[e.g.][]{combes1990,norman1996,korme2004,athanassoula2005}. These results suggest the  two-infall chemical-evolution model may represent the scenario capable of  reproducing the chemical evolution that we find in the Golden Sample.

In the context  of the  two-infall model, the adopted analytical expression  for  the gas infall rate is:

\begin{eqnarray}
\mathcal{I}_i(t)&=& \overbrace{\mathcal{X}_{1,i} \mathcal{A} \, e^{-t/ \tau_{1}}}^{\text{{{1st infall}}}}+ 
 \overbrace{\theta(t-t_{{\rm max}}) \, \mathcal{X}_{2,i} \, \mathcal{B} \, e^{-(t-t_{{\rm max}})/ \tau_{2}}}^{\text{{{2nd infall}}}},
\label{infall}
 \end{eqnarray}
where  $\tau_{1}= 0.4$ Gyr and  $\tau_{2}=2$ Gyr   are the timescales of the two distinct gas-infall episodes. 
The Heaviside step function is represented by $\theta$. 
  $\mathcal{X}_{1,i}$ and $\mathcal{X}_{2,i}$  are the abundance by mass unit of the element $i$ in the
infalling gas for the first and  second gas infall, respectively.
The quantity $t_{{\rm max}}$ is the time of the maximum infall rate on the  second accretion episode; that is, it indicates the delay  between the two peaks of 
 infall rates.
In this model, gas outflow occurs during the first infall phase ---in which the accretion of gas happens on a short timescale, and is therefore faster---  with a rate proportional to the SFR through a free parameter. The outflow rate is defined as
    \begin{equation}
    \dot\sigma_{w}=-\omega \cdot \psi(t) \, 
    \begin{cases}
      \omega=10 & \text{if  $t<t_{max}$}\\
      \omega=0 & \text{if  $t>t_{max}$},\\
     
    \end{cases}       
    \end{equation}
where    $\omega$ is the mass loading of the wind. Values of 10 or even greater for this
latter parameter are expected in galaxies with stellar masses < $10^9$ M$_{\odot}$ (e.g. \citealt{chisholm2017}). Hence, the large mass-loading factors experienced by the inner bulge in a scenario of an early, fast collapse  (first gas infall episode) can impose important  constraints on the host environment of the stars residing in the bulge (or possibly even the mass of the MW bulge when these stars formed; see analysis for MW-mass systems by  \citealt{vandokkum2013}).

  The star formation rate (SFR)  is expressed as the \citet{kenni1998} law,
\begin{equation}
\psi(t)\propto \nu \cdot \sigma_{g}(t)^{k},
\label{k1}
\end{equation}
 where $\sigma_g$ is the gas surface
 density and $k = 1.5$ is the exponent.  The quantity $\nu$ is the star formation efficiency (SFE) and this has been set to the value of 10 Gyr$^{-1}$, a value compatible with previous chemical-evolution models for the Galactic bulge \citep{matteucci2019}.
Following \citet{grieco2012}, we impose that for the second gas infall,  a chemical enrichment must be obtained from the model of the early collapse phase,  corresponding to  [Fe/H]=-1 dex.

 From   Fig. \ref{fig:final trends with model}, we note that our model predictions accurately reproduce  the Golden Sample data (NLTE) both in the [Mg/Fe] versus [Fe/H] and [Si/Fe] versus [Fe/H] relations. It is worth mentioning that, for Mg,  the agreement between data and the  model substantially improves  if we consider NLTE stars.
Nevertheless,  the proposed two-infall model underestimates the  data  in the [Ca/Fe] versus [Fe/H] space.  
 This is due to the  adopted nucleosynthesis prescriptions. In fact, we make use of the ones  suggested by \citet[][in their model 15,]{romano2010} and we refer the reader to this article for  details. We stress here that  in Fig. 22 of this latter paper, model predictions  for the Ca  in the solar neighbourhood  also  substantially underestimate that data sample in the [Ca/Fe] versus [Fe/H] space  (see also the discussion in  \citealt{spitoni2022}).

\begin{figure}[ht!]
   \centering
   \includegraphics[width=\hsize]{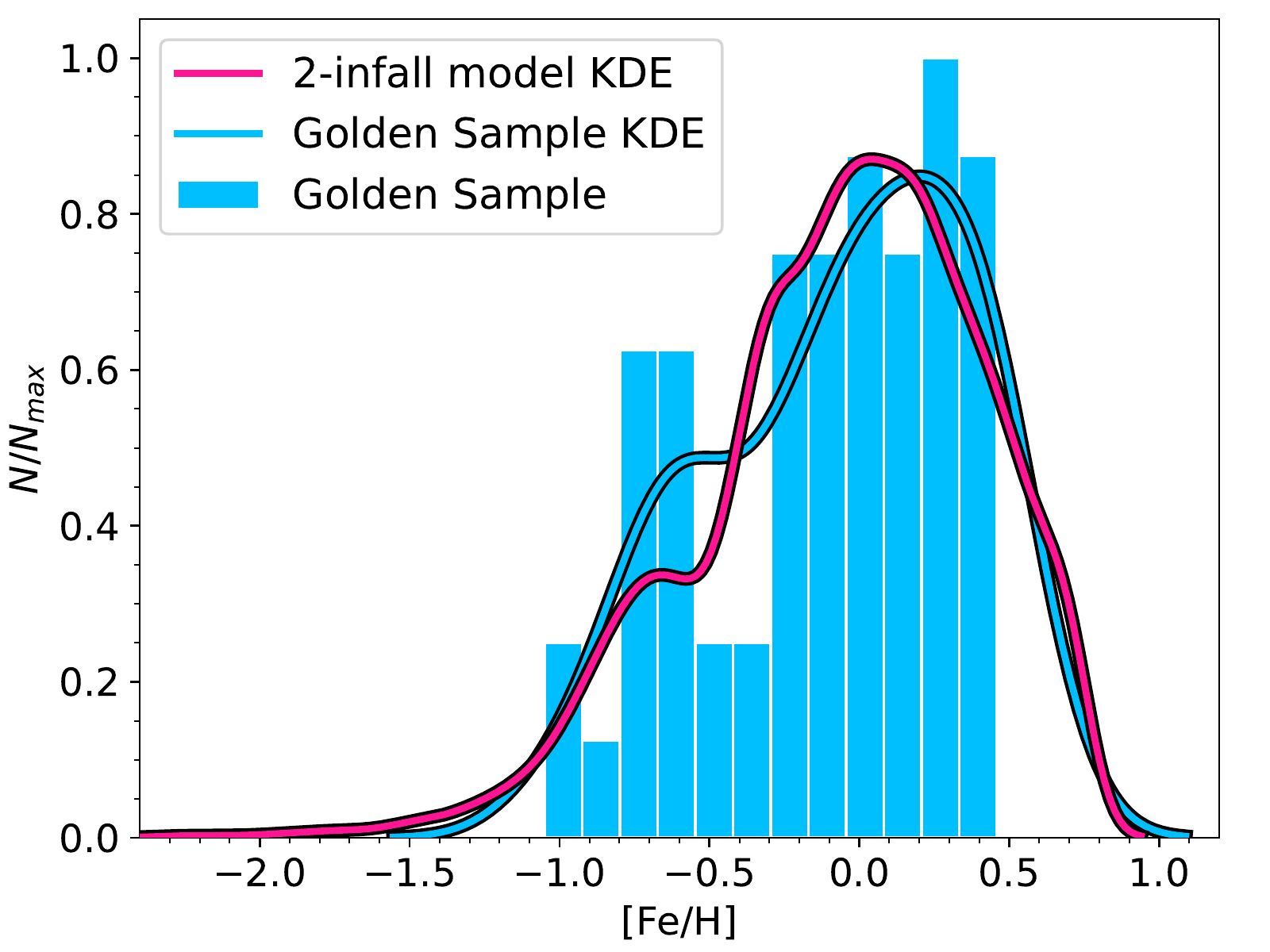}
      \caption{Histogram and kernel density estimation of the metallicity distribution function of our Golden Sample stars (blue histogram and blue line) compared to those predicted by our chemical-evolution model (pink line) presented Section \ref{sec:2-infall}.}
         \label{fig:MDF + model}
\end{figure}

The histogram and kernel density estimate (KDE) of the MDF of our Golden Sample (see Section \ref{sec:golden sample} for more details about the star selection) is shown Fig. \ref{fig:MDF + model}.
In this figure,  we can also appreciate that our two-infall model for the inner bulge  accurately reproduces the MDF of the Golden Sample.
It is important to underline that the purpose of this study is not to analyse the MDF in the inner Galactic bulge in detail, since the targeted stars have not been specifically selected using an unbiased method concerning this parameter. The stars were selected in such  a way that they are sufficiently bright and therefore do not suffer from excessive interstellar reddening but cover a wide range in the colour and therefore effective temperature. Because of our limited sample size, we do not address any corrections concerning the selection function. However, MDF studies so far (see e.g. \citealt{Rojas2020}, \citealt{Nandakumar2017}) do not reveal strong selection effects on the shape of the MDF.
Compared to Paper I, no particular offset is observed with our new metallicity values. 

\section{Conclusions}
We analysed the $\alpha$-element composition of 72 M giants in the inner Galactic bulge from \cite{nandakumar:18}  using high-resolution CRIRES spectra. To this end, we determined precise silicon, magnesium, and calcium abundances using a meticulous star-by-star analysis. In addition, thanks to a recently updated K-band line list, improved broadening parameters, a more precise macro-turbulence estimation, and NLTE corrections, this paper provides one of the most detailed studies of $\alpha$-abundance trends in the inner bulge carried out so far. This study can be used as a kind of benchmark for future spectroscopic surveys such as the upcoming MOONS galactic survey  (\citealt{MOONS}).

Thanks to a membership study using distance estimations and simulated orbits, we find that four stars of our sample are not bulge members. With additional selections based on S/N and line-by-line discrepancy, we constructed a clean sample of our inner bulge stars.

We compare our $\alpha$-abundance trends in LTE and NLTE cases and find the NLTE corrections reduce scatter overall, with Mg having the most reduction.  However, the NLTE analysis results in $\sim 0.1$ dex lower [Mg/Fe]; applying NLTE corrections gives a noticeably lower trend in [Mg/Fe] versus [Fe/H].  
 Calcium owes its higher dispersion to the lower quality of the Ca lines used. 
NLTE [$\alpha$/Fe] versus [Fe/H] trends of our inner bulge stars and those for thin/thick discs of our solar neighbourhood sample obtained using the same lines and spectral analysis show similar behaviour, with the bulge sample extending to +0.5 dex in [Fe/H], +0.3 dex beyond the solar neighborhood sample.  For both Si and Mg, the bulge trend at suprasolar metallicity is lower than the solar vicinity, especially for Mg.  This is striking and is contradictory to the canonical alpha enhancement expected for bulge giants; this trend requires confirmation.

However, for metal-rich stars with [Fe/H]>0.3, we observe a flattening in the [Si/Fe] and [Mg/Fe] thin-disc trends, which we do not find in our inner bulge trends nor in our chemical-evolution model trends. Additional solar neighbourhood  stars for these values of metallicity would be needed to validate this difference.

Based on our derived MDF and $\alpha$-abundance trends, we constructed a tailored chemical-evolution model for the inner Galactic bulge.   This latter is highly consistent with silicon and magnesium abundances. In the case of calcium, our model underestimates the abundance values because of the nucleosynthesis prescriptions selected in this work. According to this model, two gas-accretion events with timescales of 0.4 Gyr and 2 Gyr would explain the chemical fingerprints in the $\alpha$-elements of these inner-bulge stars.
However, a larger sample of high-quality abundances in the inner bulge is clearly needed in order to confront them with galactic chemical-evolution models.

\begin{acknowledgements}
We are grateful to the referee for her/his useful comments. We would like to thank the IGRINS team and especially Dr. G. Mace for excellent help and support. Further, we would like to acknowledge to  enlightening discussions with Dr. A. Amarsi about NLTE calculations.  
This work used the Immersion Grating Infrared Spectrometer (IGRINS) that was developed under a collaboration between the University of Texas at Austin and the Korea Astronomy and Space Science Institute (KASI) with the financial support of the Mt. Cuba Astronomical Foundation, of the US National Science Foundation under grants AST-1229522 and AST-1702267, of the McDonald Observatory of the University of Texas at Austin, of the Korean GMT Project of KASI, and Gemini Observatory. The paper is based on observations collected at the European Southern Observatory, Chile, program numbers 089.B-0312(A)/VM/CRIRES,
089.B-0312(B)/VM/ISAAC, 091.B-0369(A)/VM/SOFI, and 091.B-0369(B)/SM/CRIRES.
N.R. acknowledges support from the Royal Physiographic Society in Lund through the Stiftelsen Walter Gyllenbergs fond and Märta och Erik Holmbergs donation. N.R. ackowledges the support from Magnus Bergvalls stiftelse. M.S. and N.N. acknowledge the funding of the BQR Lagrange. OA and FR acknowledge financial support from the Knut and Alice Wallenberg Foundation and the Swedish Research Council (grant 2019-04659). E.S. received funding from the European Union’s Horizon 2020 research and innovation program under SPACE-H2020 grant agreement number 101004214 (EXPLORE project). PSB acknowledges support from the Swedish Research Council through an individual project grant with contract No. 2020-03404. N.N. thanks Gabriele Contursi for helpful comments on figures.
\end{acknowledgements}

\bibliography{references}{} 
\bibliographystyle{aa} 

%
%

\end{document}